\newcommand{\diracslash}[1]{#1\llap{/\kern2pt}}

\newcommand{\be}{\begin{equation}}
\newcommand{\ee}{\end{equation}}
\newcommand{\bea}{\begin{eqnarray}}
\newcommand{\eea}{\end{eqnarray}}
\newcommand{\ba}[1]{\begin{array}{#1}}
\newcommand{\ea}{\end{array}}

\newcommand{\bt}{\begin{tabular}}
\newcommand{\et}{\end{tabular}}

\newcommand{\beas}{\begin{eqnarray*}}
\newcommand{\eeas}{\end{eqnarray*}}

\documentclass[preprint,prd,aps,amssymb,amsmath
,floats,nofootinbib,floatfix]{revtex4}

\DeclareSymbolFont{rsfs}{U}{rsfs}{m}{n}
\DeclareSymbolFontAlphabet{\mathrsfs}{rsfs}

\usepackage{graphicx}
\usepackage{multirow}
\usepackage{graphicx,epstopdf}

\usepackage{graphicx}
\usepackage{float}

\usepackage[utf8]{inputenc}
\usepackage[english]{babel}
\usepackage{hyperref}
\hypersetup{
    colorlinks=true,
    linkcolor=blue,
    filecolor=magenta,     
    citecolor=blue, 
    urlcolor=blue,
}
 
\usepackage{cleveref}

\begin{document}

\title{Analysis of pseudoscalar and scalar $D$ mesons and charmonium decay width in  hot  magnetized asymmetric nuclear matter} 

\author{Rajesh Kumar}
\email{rajesh.sism@gmail.com}
\author{Arvind Kumar}
\email{iitd.arvind@gmail.com, kumara@nitj.ac.in}
\affiliation{Department of Physics, Dr B R Ambedkar National Institute of Technology Jalandhar, 
 Jalandhar -- 144011,Punjab, India}


\def\be{\begin{equation}}
\def\ee{\end{equation}}
\def\bearr{\begin{eqnarray}}
\def\eearr{\end{eqnarray}}
\def\zbf#1{{\bf {#1}}}
\def\bfm#1{\mbox{\boldmath $#1$}}
\def\hf{\frac{1}{2}}
\def\kp{\zbf k+\frac{\zbf q}{2}}
\def\km{-\zbf k+\frac{\zbf q}{2}}
\def\hwo{\hat\omega_1}
\def\hwt{\hat\omega_2}

\begin{abstract}
In this article, we calculate the mass shift and decay constant of isospin averaged pseudoscalar ($D^+$,$D^0$) and scalar ($D^+_0$,$D^0_0$) mesons by the magnetic field induced quark and gluon condensates at finite density and temperature of asymmetric nuclear matter. We have calculated the in-medium chiral condensates from the chiral SU(3) mean field model and subsequently used these condensates in QCD Sum Rules (QCDSR) to calculate the effective mass and decay constant of $D$ mesons. 
Consideration of external magnetic field effects in hot and dense nuclear  matter lead to appreciable modification in the masses and decay constants of  $D$ mesons. Furthermore, we also studied the effective decay width of higher charmonium states ($\psi(3686),\psi(3770),{{\chi_c}_0}(3414),{{\chi_c}_2}(3556)$) as a by-product by using $^3P_0$ model which can have an important impact on the yield of $J/\psi$ mesons. The results of present work will be helpful to understand the experimental observables of the heavy ion colliders which aim to produce matter at finite density and moderate temperature.

\end{abstract}

\maketitle

\section{Introduction}
\label{intro}

The construction of future Heavy Ion Collider (HIC), such as Japan Proton Accelerator Research Complex (J-PARC Japan), Compressed Baryonic Matter (CBM, GSI Germany), Proton AntiProton Annihilation in Darmstadt (PANDA, GSI Germany) and Nuclotron-based Ion Collider Facility (NICA, Dubna Russia) will shed light on the non-perturbative regime of the QCD by exploring the hadronic matter in high density and moderate temperature range \cite{Rapp2010}. In HICs, two heavy ion beams are smashed against each other and as a byproduct  Quark Gluon Plasma (QGP) comes into existence under extreme conditions of temperature and density, but it lives for a very short interval of time \cite{Vogt2007}. Subsequently, with the decrease of temperature, phase transition occurs in which QGP gets modified into hadronic matter by the process called hadronization \cite{Vogt2007}. Alongside the medium attributes such as isospin asymmetry (due to unequal no. of protons and neutrons in heavy ion), strangeness (due to the presence of strange particles in the medium), temperature and density, recently it was found that in HICs, a strong magnetic field is also produced having field strength  $eB$ $\sim$ $2-15 m_{\pi}^2$  (1${{m}_{\pi}^2}$ = $ 2.818\times 10^{18}$ gauss) approximately \cite{Kharzeev2008,Fukushima2008,Skokov2009}. Since then, physicists are trying to understand how the presence of magnetic field affects the I$^{st}$ and II$^{nd}$ order phase transitions \cite{Kharzeev2013,Fukushima2008,Vilenkin1980,
Burnier2011,Kumar2019}. The time duration for which the magnetic field remains is a very debateable topic. Many theories suggest that the magnetic field produced in HICs does not die immediately due to the interaction of itself with the medium. The primary magnetic field induces electric current in the matter and due to Lenz's law, a secondary magnetic field comes into picture which slows down the decay rate of the magnetic field \cite{Tuchin2011,Tuchin2011a,Tuchin2013,Marasinghe2011,Das2017,
Reddy2018,Cho2015,Kumar2019}. These interactions increase the electric conductivity of the medium which further affects the relaxation time of the magnetic interaction and this phenomenon is called the chiral magnetic effect \cite{Kharzeev2013,Fukushima2008,Vilenkin1980,
Burnier2011}. The presence of magnetic field affects the yield of in-medium/vacuum chiral condensates hence the location of critical temperature $T_c$ is also affected and this process is known as (inverse) magnetic catalysis \cite{Kharzeev2013}.

 Near the hadron phase transition, it is not possible to detect QGP directly due to its short-lived nature, hence many other indirect observations are used as a tool to understand its existence namely jet quenching \cite{Bjorken1982}, strangeness enhancements \cite{Soff1999,Capella1995}, dilepton enhancements \cite{Masera1995,Srivastava2009,Wilson1998} and $J/\psi$ ($\Upsilon$) suppression \cite{Matsui1986}. In 1986, Matsui and Satz proposed the idea of $J/\psi$ suppression on the basis of color debye screening \cite{Matsui1986}. In this mechanism, when the debye screening radius becomes less than the charm quark system's binding radius, the charm binding force can no longer keep $c$ and $\bar c$ quark together. These free charm quarks (antiquarks)  form bound state with free light quarks ($u,d,s$) in the medium to form  $D$ mesons. The in-medium effects on open charm mesons are more than the quarkonium. This is due to the fact that the in-medium properties of $D$ mesons depend upon light quark condensates which varies appreciably with the medium whereas, for charmonia (bottomonia), it depends on gluon condensates which do not change much with density \cite{Kumar2019,Chhabra2018}. It may also be noted that $J/\psi$ suppression occurs not only due to QGP formation, but because of, density-dependent suppression, comover scattering \cite{Gerschel1988} and nuclear dependence of $D$ and $B$ (for $\Upsilon$ suppression) mesons \cite{Garcia2011,Zhang2000} too. Higher bottomonium and charmonium states decay to  $\Upsilon$ and $J/\psi$ mesons respectively and hence are considered to be the major source of these ground state mesons \cite{Chhabra2017,Matsui1986}. Under the effect of different medium conditions, if the mass of $D$ ($B$) meson  decrease appreciably, then these higher quarkonium state will prefer to decay in $D \bar D$ ($B \bar B$) meson pair rather decaying in conventional $J/\psi$ ($\Upsilon$) meson. In $AA$ and $p\bar A$ collisions, the decay width  of higher quarkonium states and other experimental observables \cite{Inghirami2019} can be directly measured experimentally to validate the phenomenological results \cite{Friman2002}.
 
 QCD Phase diagram is a graphical  representation  to account the different QCD regime's physics with different medium parameters. To study  this diagram in hadron phase, several potential models are constructed on the basis of effective field theory by incorporating basic properties of QCD notably broken scale invariance and symmetry breaking \cite{Papazoglou1999,Kumar2019}. Some of these models are:   Walecka
model \cite{Walecka1974}, Nambu-Jona-Lasinio (NJL) model \cite{Nambu1961},   chiral SU(3) model \cite{Papazoglou1999,Mishra2004a,Mishra2009,Kumar2010,
Kumar2019}, QCD sum rules  \cite{Reinders1981,Hayashigaki2000,
Hilger2009,Reinders1985,Klingl1997,Klingl1999}, Quark-Meson Coupling (QMC) model \cite{Guichon1988,Hong2001,Tsushima1999,Sibirtsev1999,Saito1994,Panda1997},    and  coupled channel approach \cite{Tolos2004,Tolos2006,Tolos2008,Hofmann2005}. In  above approaches, the effect of thermal and quantum fluctuations are neglected by using  mean field approximations. These fluctuations are included by modified potential models such as  Polyakov Quark Meson (PQM) model \cite{Chatterjee2012,Schaefer2010}, the Polyakov loop extended NJL  (PNJL) model \cite{Fukushima2004,Kashiwa2008,Ghosh2015} and Functional Remormalization Group (FRG) \cite{Herbst2014,Drews2013} techniques. In addition, the  decay width of the heavy mesons have  been explained through various models, $i.e.,$  $^3 S_1$ model \cite{Furui1987}, elementary meson-emission model \cite{Bonnaz1999}, flux-tube model \cite{Kokoski1987} and $^3 P_0$ model \cite{Friman2002}. 

Our present work is in a threefold way. At first, we calculate the in-medium quark and gluon condensates from the chiral SU(3) mean field model and secondly use them in QCDSR to calculate the medium induced mass and decay constant of $D$ mesons in the presence of magnetic field. At last, by using $^3P_0$ model, we study the magnetic field induced decay width of higher charmonium states. The in-medium properties of  meson under the effect of strong magnetic fields have been studied by various non-perturbative techniques in the literature \cite{Wang2015,Wang2011,Kumar2014,Chhabra2017}.
For example, in Ref. \cite{Reddy2018}, the properties of $D$ meson  in strongly magnetized asymmetric nuclear matter was studied using chiral SU(4) model and observed an additional  positive mass shift  for the charged $D$ meson, due to interaction with the magnetic field. In addition to this, under the effect of magnetic field the mass spectra of $D$ mesons  and mixing effects between pseudoscalar and vector $D$ mesons have been studied with the use of Operator Product Expansion technique of QCDSR by Gubler $et. al.$ \cite{Gubler2016}. The magnetic induced decay width of higher charmonium states into lower charmonium states are calculated with the joint approach of chiral model and $^3P_0$ model \cite{Mishra2019}. In this work, author observed appreciable magnetic field effects in the cold nuclear matter . The process such as chiral magnetic effect and (inverse) magnetic catalysis shows great effect on the physics of deconfinement and chiral symmetry breaking. The analytic crossover, critical point and phase transition of QCD Phase diagram is studied extensively in the literature \cite{Bali2012,Kharzeev2013,Gatto2011}.  In addition to these articles, the effect of strong magnetic fields is also studied on the properties of $\rho$ meson \cite{Liu2015}, $B$ meson \cite{Machado2014,Dhale2018}, charmonium \cite{Cho2015,Cho2014,Kumar2019,Kumar2019a} and bottomonium states \cite{Kumar2019a,Jahan2018}. A lot of work has also been done without taking the effect of magnetic field. For example,  Tolos $et. al.$ investigated the increase in the mass of $D$ mesons in the nuclear medium using a coupled-channel approach \cite{Tolos2008}. In the QMC model, Tsushima and Khanna  observed a negative shift of  $D$ mesons in the nuclear medium and  also discussed the possibility of the formation of $D$ mesic nuclei due to the attractive interaction of $D$ meson with the medium constituents \cite{Tsushima2003}. The chiral SU(3) model was generalised to SU(4) sector to study the in-medium mass of pseudoscalar $D$ mesons \cite{Kumar2011}. In this article, along with the in-medium mass of $D$ meson, authors also studied the decay width of higher charmonium states into $D \bar D$ pairs using $^3P_0$ model. Using QCD sum rules, Wang $et. al.$  calculated the mass and decay constant of pseudoscalar, scalar, vector and axial vector $D$ mesons by taking the contributions from next to leading order terms \cite{Wang2015}. Using QCDSR, the contribution up to leading order term have also been used to calculate the properties of scalar $D$ mesons \cite{Hayashigaki2000}.  By using the unification of chiral SU(3) model and QCDSR, the in-medium mass and decay constant of pseudoscalar, scalar, vector and axial vector $D$ mesons are calculated in the strange hadronic medium and observed a negative (scalar and vector) and positive shift (pseudo scalar and axial vector) in the mass of $D$ mesons \cite{Kumar2015,Chhabra2017,Chhabra2018,Kumar2014}. In these articles, authors have also calculated the in-medium decay width of higher charmonium states \cite{Chhabra2017} and scalar $D$ mesons \cite{Chhabra2018}. The in-medium decay width of different heavy charmonia is also calculated using quark anti-quark pair $^3P_0$ model \cite{Friman2002} and recently this model was also used to calculate the decay width of $\psi$(4260) \cite{Bruschini2019}.

 The outline of the present paper is as follows: In the forthcoming subsection \ref{subsec:2.1} and \ref{subsec:2.2}, we will briefly explain the formalism to calculate the effective masses and decay constant of pseudoscalar and scalar $D$ mesons under the effect of magnetic field. In subsection \ref{subsec:2.3}, we will describe the methodology to calculate the decay width of higher charmonium states. In section \ref{sec:3}, we will discuss the quantitative results of the present work and at last in section \ref{sec:4}, we will give a conclusion.

\section{ Formalism}
\label{sec:2}

 We use the unification of chiral SU(3) model and QCDSR techniques to study the effective mass and decay constant of scalar and pseudoscalar $D$ mesons. These non pertutbative techniques are constructed to understand the low energy QCD by using renormalization methods\cite{Papazoglou1999,Kumar2010,Hayashigaki2000,Reinders1985}. In this section, we gradually discuss the quark and gluon condensates, in-medium mass and decay constant of $D$ mesons and the  magnetic field induced charmonium decay width.
\
 \subsection{Quark and Gluon Condensates from Chiral SU(3) Model }
 \label{subsec:2.1}

We use the non-perturbative chiral SU(3) model, constructed on the basis of effective field theory. This model incorporates the basic QCD features such as non-linear realization of chiral symmetry and trace anomaly\cite{Weinberg1968,Coleman1969,Zschiesche1997,Bardeen1969,
Kumar2010,Papazoglou1999,Kumar2019}. In this framework, the trace anomaly (broken scale invariance) property of QCD is preserved by the introduction of scalar dilaton field $\chi$\cite{Papazoglou1999,Kumar2019}. Also, the isospin asymmetry of the medium is incorporated by the introduction of scalar isovector delta field $\delta$ and vector-isovector field $\rho$ \cite{Kumar2010}. The model is built under the assumption of mean field potential in which the mixing of vector and pseudoscalar mesons have been neglected, hence the effect of thermal and quantum fluctuations are not studied in the present work\cite{Kumar2010,Reddy2018}. The effect of the external magnetic field is taken by adding the Lagrangian density due to magnetic field in the chiral effective Lagrangian density\cite{Kumar2019,Reddy2018}.  By minimizing the thermodynamic potential of chiral SU(3) model\cite{Kumar2019,Kumar2019a}, the coupled equations of motion  of the  scalar ($\sigma$, $\zeta$, $\delta$, $\chi$), and vector ($\omega$,$\rho$), meson exchange fields are derived and are given as

\begin{eqnarray}
 k_{0}\chi^{2}\sigma-4k_{1}\left( \sigma^{2}+\zeta^{2}
+\delta^{2}\right)\sigma-2k_{2}\left( \sigma^{3}+3\sigma\delta^{2}\right)
-2k_{3}\chi\sigma\zeta \nonumber\\
-\frac{d}{3} \chi^{4} \bigg (\frac{2\sigma}{\sigma^{2}-\delta^{2}}\bigg )
+\left( \frac{\chi}{\chi_{0}}\right) ^{2}m_{\pi}^{2}f_{\pi}
=\sum g_{\sigma i}\rho_{i}^{s} ,
\label{sigma}
\end{eqnarray}
\begin{eqnarray}
 k_{0}\chi^{2}\zeta-4k_{1}\left( \sigma^{2}+\zeta^{2}+\delta^{2}\right)
\zeta-4k_{2}\zeta^{3}-k_{3}\chi\left( \sigma^{2}-\delta^{2}\right)\nonumber\\
-\frac{d}{3}\frac{\chi^{4}}{\zeta}+\left(\frac{\chi}{\chi_{0}} \right)
^{2}\left[ \sqrt{2}m_{K}^{2}f_{K}-\frac{1}{\sqrt{2}} m_{\pi}^{2}f_{\pi}\right]
 =\sum g_{\zeta i}\rho_{i}^{s} ,
\label{zeta}
\end{eqnarray}
\begin{eqnarray}
k_{0}\chi^{2}\delta-4k_{1}\left( \sigma^{2}+\zeta^{2}+\delta^{2}\right)
\delta-2k_{2}\left( \delta^{3}+3\sigma^{2}\delta\right) +2k_{3}\chi\delta
\zeta \nonumber\\
 +   \frac{2}{3} d \chi^4 \left( \frac{\delta}{\sigma^{2}-\delta^{2}}\right)
=\sum g_{\delta i}\tau_3\rho_{i}^{s}  ,
\label{delta}
\end{eqnarray}

\begin{eqnarray}
k_{0}\chi \left( \sigma^{2}+\zeta^{2}+\delta^{2}\right)-k_{3}
\left( \sigma^{2}-\delta^{2}\right)\zeta + \chi^{3}\left[1
+{\rm {ln}}\left( \frac{\chi^{4}}{\chi_{0}^{4}}\right)  \right]
+(4k_{4}-d)\chi^{3}
\nonumber\\
-\frac{4}{3} d \chi^{3} {\rm {ln}} \Bigg ( \bigg (\frac{\left( \sigma^{2}
-\delta^{2}\right) \zeta}{\sigma_{0}^{2}\zeta_{0}} \bigg )
\bigg (\frac{\chi}{\chi_0}\bigg)^3 \Bigg )+
\frac{2\chi}{\chi_{0}^{2}}\left[ m_{\pi}^{2}
f_{\pi}\sigma +\left(\sqrt{2}m_{K}^{2}f_{K}-\frac{1}{\sqrt{2}}
m_{\pi}^{2}f_{\pi} \right) \zeta\right] \nonumber\\
-\frac{\chi}{{{\chi_0}^2}}(m_{\omega}^{2} \omega^2+m_{\rho}^{2}\rho^2)  = 0 ,
\label{chi}
\end{eqnarray}

\begin{eqnarray}
\left (\frac{\chi}{\chi_{0}}\right) ^{2}m_{\omega}^{2}\omega+g_{4}\left(4{\omega}^{3}+12{\rho}^2{\omega}\right) =\sum g_{\omega i}\rho_{i}^{v}  ,
\label{omega}
\end{eqnarray}

and

\begin{eqnarray}
\left (\frac{\chi}{\chi_{0}}\right) ^{2}m_{\rho}^{2}\rho+g_{4}\left(4{\rho}^{3}+12{\omega}^2{\rho}\right)=\sum g_{\rho i}\tau_3\rho_{i}^{v}  ,
\label{rho}
\end{eqnarray}

respectively.

In above, the parameters $k_0, k_2$ and $k_4$ are fitted so as to reproduce the vacuum values of scalar meson fields and the other parameters, such as  $k_1$ is constrained to obtain the in-medium mass of nucleon at nuclear saturation density, $\rho_N$ and the parameter $k_3$ is selected so as to generate the masses of $\eta$ and   $\eta^\prime$  mesons. In addition, the parameters $f_\pi$, $f_K$ and $m_\pi$, $m_K$  are the decay constants and masses of pions and kaons, respectively. Moreover, the effect of isospin asymmetry is introduced in the nuclear matter calculations by the parameter ($\eta = -\frac{\Sigma_i \tau_{3i} \rho^{v}_{i}}{2\rho_{N}}$). Where $\rho^{s}_{i}$ and $\rho^{v}_{i}$ represent the scalar and vector densities of $i^{th}$ nucleon ($i=n,p$)  in the presence of magnetic field which is applied in the $Z$-direction \cite{Kumar2019,Broderick2000,Broderick2002} and $\tau_{3i}$ is the $I_3$ component of isospin. With the interaction of protons with magnetic field, the Landau quantization takes place\cite{Broderick2000,Kumar2019}. This circular confined motion disrupt the net momentum $k$ in two parts, $i.e.,$ $k_{\bot}$ (perpendicular to the $Z$-axis) and $k_{\parallel}$ (parallel to $Z$-axis)\cite{Kumar2019}.

The magnetic field induced  scalar density as well as
the vector density of uncharged neutron given in Eqs.(\ref{sigma}) and (\ref{omega})  are given as \cite{Broderick2000,Broderick2002}

\bea
\rho^{s}_{n}&=&\frac{1}{2\pi^{2}}\sum_{s=\pm 1}\int^{\infty}_{0}
k^{n}_{\bot}dk^{n}_{\bot}\left(1-\frac{s\mu_{N}\kappa_{n}B}
{\sqrt{m^{* 2}_{n}+\left(k^{n}_{\bot}\right)^{2}}} \right)  
\int^{\infty}_{0}\, dk^{n}_{\parallel}
\frac{m^*_n}{\tilde E^{n}_{s}}\left(
f^n_{k, s}+\bar{f}^n_{k, s}\right),
\label{rhosn} 
\eea
and
\bea
\rho^{v}_{n}&=&\frac{1}{2\pi^{2}}\sum_{s=\pm 1}\int^{\infty}_{0}k^{n}_{\bot}
dk^{n}_{\bot} \int^{\infty}_{0}\, dk^{n}_{\parallel}
\left( f^n_{k, s}-\bar{f}^n_{k, s}\right), 
\label{rhovn} 
\eea

respectively. Similarly, for the charged proton, the scalar  and vector densities are given by \cite{Broderick2000,Broderick2002}

\begin{equation}
\rho^{s}_{p}=\frac{|q_{p}|Bm^{*}_{p}}{2\pi^2} \Bigg [ 
\sum_{\nu=0}^{\nu_{max}^{(s=1)}}\int^{\infty}_{0}\frac{dk^p_{\parallel}}{\sqrt{(k^{p}_{\parallel})^2
+(\bar m_{p})^2}}\left( f^p_{k,\nu, s}+\bar{f}^p_{k, \nu, s}\right)
+\sum_{\nu=1}^{\nu_{max}^{(s=-1)}} \int^{\infty}_{0}\frac{dk^p_{\parallel}}{\sqrt{(k^{p}_{\parallel})^2
+(\bar m_{p})^2}}\left( f^p_{k,\nu, s}+\bar{f}^p_{k, \nu, s}\right)
\Bigg],
\label{rhosp}
\end{equation}
and

\begin{eqnarray}
\rho^{v}_{p}=\frac{|q_{p}|B}{2\pi^2} \Bigg [ 
\sum_{\nu=0}^{\nu_{max}^{(s=1)}} \int^{\infty}_{0}
dk^p_{\parallel}\left( f^p_{k,\nu, s}-\bar{f}^p_{k,\nu, s}\right)
+\sum_{\nu=1}^{\nu_{max}^{(s=-1)}} \int^{\infty}_{0}
dk^p_{\parallel}\left( f^p_{k,\nu, s}-\bar{f}^p_{k,\nu, s}\right) 
\Bigg],
\label{rhovp}
\end{eqnarray}

respectively, where $\bar m_{p}$ is the induced mass under the effect of magnetic field, which is defined as
\bea
\bar m_{p}=\sqrt{m^{* 2}_{p}+2\nu |q_{p}|B}-s\mu_{N}\kappa_{p}B.
\label{mc}
\eea

In above equations, $\nu$ represents the Landau quantized levels and $m_{i}^{*} = -(g_{\sigma i}\sigma + g_{\zeta i}\zeta + g_{\delta i}\tau_{3i} \delta)$ is the effective mass of the nucleons. Here, $g_{\sigma i}$, $g_{\zeta i}$ and $g_{\delta i}$ represent the coupling constants of  $i^{th}$ nucleons with  $\sigma$, $\zeta$ and $\delta$ fields respectively.  The effective single particle energy of proton is given by $\tilde E^{p}_{\nu, s}=\sqrt{\left(k^{p}_{\parallel}\right)^{2}+
\left(\sqrt{m^{* 2}_{p}+2\nu |q_{p}|B}-s\mu_{N}\kappa_{p}B \right)^{2}}$, whereas for neutron its expression  is given as $\tilde E^{n}_{s}= \sqrt{\left(k^{n}_{\parallel}\right)^{2} +
\left(\sqrt{m^{* 2}_{n}+\left(k^{n}_{\bot}\right)^{2} }-s\mu_{N}\kappa_{n}B 
\right)^{2}}$. The constants $k_i$ and $s$ are the anomalous magnetic moment and the spin of the nucleons respectively.
 In addition, ${f}^n_{k, \nu, s}$, $\bar{f}^n_{k, \nu, s}$,  ${f}^p_{k, s}$ and $\bar{f}^p_{k, s}$ represent the  finite temperature distribution functions   for neutron and proton and their antiparticles, and are given as
\bea
f^n_{k, s} &=& \frac{1}{1+\exp\left[\beta(\tilde E^n_{s} 
-\mu^{*}_{n}) \right]}, \qquad
\bar{f}^n_{k, s} = \frac{1}{1+\exp\left[\beta(\tilde E^n_{s} 
+\mu^{*}_{n} )\right]}.
\label{dfn}
\eea
\bea
f^p_{k,\nu, s} &=& \frac{1}{1+\exp\left[\beta(\tilde E^p_{\nu, s} 
-\mu^{*}_{p}) \right]}, \qquad
\bar{f}^p_{k,\nu, s} = \frac{1}{1+\exp\left[\beta(\tilde E^p_{\nu, s} 
+\mu^{*}_{p} )\right]}.
\label{dfp}
\eea

As we will see later, to calculate the effective mass and decay constant of scalar and pseudo scalar $D$ mesons using QCDSR, we need the values of quarks and gluon condensates. In chiral model, the scalar quark condensates  can be related to symmetry breaking via relation\cite{Kumar2019}
\begin{equation}
\sum_{i} m_{i}\langle \bar{q}_{i}q_{i}\rangle_{\rho_N}=-\mathcal{L}_{SB},
\label{eqsb}
\end{equation}
where $\mathcal{L}_{SB}$ is a explicit symmetry breaking Lagrangian term \cite{Kumar2019} and by using this equation we formulated the up and down quark condensates, which are expressed as


\begin{align}
\left\langle \bar{u}u\right\rangle_{\rho_N}
= \frac{1}{m_{u}}\left( \frac {\chi}{\chi_{0}}\right)^{2}
\left[ \frac{1}{2} m_{\pi}^{2}
f_{\pi} \left( \sigma + \delta \right) \right],
\label{qu}
\end{align}

and

\begin{align}
\left\langle \bar{d}d\right\rangle_{\rho_N}
= \frac{1}{m_{d}}\left( \frac {\chi}{\chi_{0}}\right)^{2}
\left[ \frac{1}{2} m_{\pi}^{2}
f_{\pi} \left( \sigma - \delta \right) \right],
\label{qd}
\end{align}

respectively. In above $m_u$ and $m_d$ are the mass of up quark and  down quark respectively.
Also, by using the broken scale invariance property of QCD \cite{Papazoglou1999,Kumar2010,Kumar2019}, the scalar gluon condensate ${G_0}_{\rho_N}$=$\left\langle \frac{\alpha_{s}}{\pi} 
G^a_{\mu \nu} {G^a}^{\mu \nu} \right\rangle_{\rho_N} $ is formulated by the comparison of energy-momentum tensor (EMT) of QCD with the EMT of chiral model and is  expressed  in terms of  scalar fields through the  relation \cite{Kumar2019}

\begin{eqnarray}
\left\langle \frac{\alpha_{s}}{\pi} 
G^a_{\mu \nu} {G^a}^{\mu \nu} \right\rangle_{\rho_N} =  \frac{8}{9} \Bigg [(1 - d) \chi^{4}
+\left( \frac {\chi}{\chi_{0}}\right)^{2} 
\left( m_{\pi}^{2} f_{\pi} \sigma
+ \big( \sqrt {2} m_{K}^{2}f_{K} - \frac {1}{\sqrt {2}} 
m_{\pi}^{2} f_{\pi} \big) \zeta \right) \Bigg ].
\label{chiglum}
\end{eqnarray}

The value of $d$=$\frac{2}{11}$  has been taken from QCD beta function, $\beta_{QCD}$  at the one loop level \cite{Papazoglou1999}.

\subsection{ Masses and Decay Constant of $D$ mesons from QCDSR}
\label{subsec:2.2}

In this subsection, we discuss the QCDSR to calculate the in-medium mass shift and decay constant of  isospin averaged pseudoscalar ($D^+,D_0$) and scalar ($D^+_0$,$D^0_0$) mesons. QCDSR is a non-perturbative technique, which is based on the Borel Transformation and Operator Product Expansion (OPE) method\cite{Reinders1985,Klingl1999,Wang2015,Hayashigaki2000}. These methods are used to deal with the divergence occurred in the asymptotic perturbative series\cite{Hayashigaki2000,Reinders1985}. We will see further the mass and decay constant of these open charm mesons is expressed in terms of scalar and gluon condensates, which contains the effect of medium parameters such as temperature, density, asymmetry and magnetic field. We start with two point current correlation function $\Pi (q)$ which represents the Fourier transformation of the time ordered product of the isospin averaged meson current, $J^{'}(x)$ and can be written as \cite{Wang2015,Chhabra2017}

\begin{align} \label{pb2}
\Pi(q) = i\int d^{4}x\ e^{iq. x} \langle \mathcal{T}\left\{J^{'}(x)J^{'\dag}(0)\right\} \rangle_{\rho_N,T},
\end{align}
where $q$ is the four momentum and $\rho_N$ and $T$ represent the nucleon density and temperature of the medium. In Ref. \cite{Hilger2009,Suzuki2016}, the mass splitting of the  different oppositely charged  mesons are also investigated  by dividing the current correlation function in even and odd terms. In this article, we have considered the average meson currents of the particle $D$ and their antiparticle $\bar D$.  The average current of scalar and pseudo scalar meson is given by the following mathematical relations
\begin{eqnarray}
 J(x) &=&J^\dag(x) =\frac{\bar{c}(x) q(x)+\bar{q}(x) c(x)}{2}\, 
 \label{mcq1}
 \end{eqnarray}
 
 and
 
  \begin{eqnarray}
  J_{5}(x) &=&J_{5}^\dag(x) =\frac{\bar{c}(x)i \gamma_5q(x)+\bar{q}(x)i\gamma_5 c(x)}{2}\,,\nonumber\\
  \label{mcq2}
\end{eqnarray}

 respectively. In above, the quark operator $q(x)$ is for $u$ and $d$ quarks and $c(x)$ is the charm quark operator. The  selection of $q$ depends upon the quark content of the given meson. From the quark composition of $D$ mesons one can easily understand that the $(D^+,D^0)$ and $(D^+_0,D^0_0)$  form the isospin doublets and they show mass splitting in the presence of isospin asymmetric medium \cite{Chhabra2017}.  Now, for the nuclear matter in the Fermi gas approximation, we divide the correlation function $\Pi(q)$ into vacuum, static nucleon and thermal part as

\begin{align}
\Pi (q) =\Pi_{0} (q)+ \frac{\rho_{N}}{2m_N}T_{N} (q) + \Pi_{P.B.}(q,T)\,,
\label{pb}
 \end{align}

 where $T_N (q)$ is the forward scattering amplitude and $m_N$ denotes the nucleon mass. The third term,$i.e.$, pion bath contribution represents the thermal effects of the medium and is given as \cite{Eletsky1995}
\begin{align} \label{pb2}
\Pi_{P.B.}(q, T) = i\int d^{4}x\ e^{iq. x} \langle \mathcal{T}\left\{J^{'}(x)J^{'\dag}(0)\right\} \rangle_{T}.
\end{align}

In the present investigation, we can neglect this thermal effects term as the temperature and magnetic effects of the medium are incorporated by quark and gluon condensates, which are calculated in terms of the meson exchange fields as discussed earlier in the subsection \ref{subsec:2.1} \cite{Kumar2015,Chhabra2018}. By neglecting the third term from Eq.(\ref{pb}), the expression becomes

\begin{eqnarray}
\Pi(q) =\Pi_{0}(q)+ \frac{\rho_N}{2m_N}T_{N}(q)\, . \nonumber \\
 \end{eqnarray}
 
The forward scattering amplitude $T_{N}(q)$  can be written as
\begin{eqnarray}
T_{N}(\omega,\mbox{\boldmath $q$}\,) &=&i\int d^{4}x e^{iq\cdot x}\langle N(p)|
T\left\{J(x)J^{\dag}(0)\right\} |N(p) \rangle\, .
\end{eqnarray}

 The amplitude $T_{N}(\omega,\mbox{\boldmath $q$}\,)$ can be related to the $DN$($D_0N$)
scattering $T$-matrix in the limit of $\mbox{\boldmath $q$}\rightarrow {\bf 0}$
\begin{eqnarray}
{{\cal T}_{D/D_0\,N}}(m_{D/D_0},0) =
8\pi(m_N+m_{D/D_0})a_{D/D_0} \, ,
\end{eqnarray}
 where   $a_{D/D_0}$ is the scattering lengths of $DN$ ($D_0N$) interactions. This scattering matrix can be represented in terms of phenomenological spectral density   $\rho(\omega,0)$, which can be parametrized in three unknown  parameters $a,\,b$ and $c$ \cite{Hayashigaki2000},
\begin{eqnarray}
\rho(\omega,0) &=& -\frac{1}{\pi} \mbox{Im} \left[\frac{{{\cal T}_{D/D_0N}}(\omega,{\bf 0})}{\left(\omega^{2}-
m_{D/D_0}^2+i\varepsilon\right)^{2}} \right]\frac{f_{D/D_0}^2m_{D/D_0}^4}{m_c^2}+ \cdots \,, \nonumber \\
&=& a\,\frac{d}{d\omega^2}\delta\left(\omega^{2}-m_{D/D_0}^2\right) +
b\,\delta\left(\omega^{2}-m_{D/D_0}^2\right) + c\,\delta\left(\omega^{2}-s_{0}\right)\, .
\label{rhow}
\end{eqnarray}
In above equation, the first term represents the double pole term, which related to the on shell effect of $T$ matrices  and can be related to the scattering length as 

\begin{eqnarray}
a_{D/D_0}=\frac{a m_c^2}{f_{D/D_0}^2m_{D/D_0}^4(-8\pi(m_N+m_{D/D_0}))}.
\label{slength}
\end{eqnarray}

Furthermore, the second term in Eq.(\ref{rhow}), represents the single pole term that relates the off-shell effects of $T$ matrices; the last term corresponds to the remaining contributions (continuum), where $s_0$ denotes the continuum threshold parameter.

The  shift in above mentioned masses $m_{D/D_0}$ and decay constant $f_{D/D_0}$   of the open charm mesons  can be written as \cite{Wang2015}
\begin{eqnarray}
\Delta m^*_{D/D_0} &=&2\pi\frac{m_{N}+m_{D/D_0}}{m_Nm_{D/D_0}}\rho_N a_{D/D_0}\, ,
\label{msd}
\end{eqnarray}

and
\begin{eqnarray}
\Delta f^*_{D/D_0}&=&\frac{m_c^2}{2f_{D/D_0}m_{D/D_0}^4}\left(\frac{b\rho_N}{2m_N}-\frac{4f_{D/D_0}^2m_{D/D_0}^3\Delta m_{D/D_0}}{m_c^2} \right) \, , \nonumber\\.
\label{fd}
\end{eqnarray}

respectively. 
Hence the effective mass of open charm mesons can be written as 
\begin{equation}
m^{*}_{D/D_0}=m_{D/D_0}+\Delta m^*_{D/D_0}.
\end{equation} 

Note that $m_{D/D_0}$ denotes vacuum mass of pseudoscalar and scalar $D$ mesons.

 As discussed earlier, the Landau quantization takes place with the interaction of charged particle with magnetic field. This interaction invokes an additional positive shift in the mass of  charged $D^{+},D^+_0$ meson and this lead to

\begin{equation}
m^{**}_{D^+/D^+_0}=\sqrt {{m^*_{D^+/D^+_0}}^2 +|eB|}.
\label{mdpm_landau}
\end{equation}
On the other hand, due to their uncharged nature the neutral pseudoscalar ($D^0$) and scalar ($D^0_0$) mesons have no modification due to magnetic field.

We get the analytic QCDSR in terms of two unknown parameters $a$ and $b$ by equating the Borel transformed forward scattering amplitude $T_{N}(\omega,\mbox{\boldmath $q$}\,)$ in the OPE side with the Borel transformed forward scattering amplitude $T_{N}(\omega,\mbox{\boldmath $q$}\,)$ in the phenomenological side \cite{Hayashigaki2000}. The parametrised QCDSR are given by  equation
\begin{eqnarray}
 a\, C_a+b\, C_b &=&C_f \, .
 \label{wdiff}
\end{eqnarray}
The explicit form of Borel transformed coefficients having  next to leading order contributions for the  pseudoscalar current $J_5(x)$ is  \cite{Wang2015},
\begin{eqnarray}
C_a &=&\frac{1}{M^2}\exp\left(-\frac{m_{D}^2}{M^2}\right)-\frac{s_0}{m_{D}^4}\exp\left(-\frac{s_0}{M^2}\right) \, ,\nonumber\\
C_b&=&\exp\left(-\frac{m_{D}^2}{M^2}\right)-\frac{s_0}{m_{D}^2}\exp\left(-\frac{s_0}{M^2}\right) \, ,
\end{eqnarray}

and 
\begin{eqnarray}
C_f&=& \frac{2m_N(m_H+m_N)}{(m_H+m_N)^2-m_{D}^2}\left(\frac{f_{D}m_{D}^2g_{DNH}}{m_c}\right)^2\left\{ \left[\frac{1}{M^2}-\frac{1}{m_{D}^2-(m_H+m_N)^2}\right] \exp\left(-\frac{m_{D}^2}{M^2}\right)\right.\nonumber\\
&&\left.+\frac{1}{(m_H+m_N)^2-m_{D}^2}\exp\left(-\frac{(m_H+m_N)^2}{M^2}\right)\right\}-\frac{m_c\langle\bar{q}q\rangle_N}{2}\left\{1+\frac{\alpha_s}{\pi} \left[ 6-\frac{4m_c^2}{3M^2} \right.\right.\nonumber\\
&&\left.\left.-\frac{2}{3}\left( 1-\frac{m_c^2}{M^2}\right)\log\frac{m_c^2}{\mu^2}-2\Gamma\left(0,\frac{m_c^2}{M^2}\right)\exp\left( \frac{m_c^2}{M^2}\right) \right]\right\}\exp\left(- \frac{m_c^2}{M^2}\right) \nonumber\\
&&+\frac{1}{2}\left\{-2\left(1-\frac{m_c^2}{M^2}\right)\langle q^\dag i Dq\rangle_N +\frac{4m_c
}{M^2}\left(1-\frac{m_c^2}{2M^2}\right)\langle \bar{q} i D i Dq\rangle_N+\frac{1}{12}\langle\frac{\alpha_sGG}{\pi}\rangle_N\right\} \nonumber\\
&&\exp\left(- \frac{m_c^2}{M^2}\right)\,.
\end{eqnarray}
 
For the scalar current  $J(x)$, we have \cite{Wang2011,Kumar2014},

\begin{eqnarray}
C_a &=&\frac{1}{M^2}\exp\left(-\frac{m_{D_0}^2}{M^2}\right)-\frac{s_0}{m_{D_0}^4}\exp\left(-\frac{s_0}{M^2}\right) \, ,\nonumber\\
C_b&=&\exp\left(-\frac{m_{D_0}^2}{M^2}\right)-\frac{s_0}{m_{D_0}^2}\exp\left(-\frac{s_0}{M^2}\right) \, ,
\end{eqnarray}
and

\begin{eqnarray}
C_f&=& \frac{2m_N(m_H-m_N)}{(m_H-m_N)^2-m_{D_0}^2}\left(\frac{f_{D_0}m_{D_0}^2g_{D_0NH}}{m_c}\right)^2\left\{ \left[\frac{1}{M^2}-\frac{1}{m_{D_0}^2-(m_H-m_N)^2}\right] \exp\left(-\frac{m_{D_0}^2}{M^2}\right)\right.\nonumber\\
&&\left.+\frac{1}{(m_H-m_N)^2-m_{D_0}^2}\exp\left(-\frac{(m_H-m_N)^2}{M^2}\right)\right\}-\frac{m_c\langle\bar{q}q\rangle_N}{2}\exp\left(- \frac{m_c^2}{M^2}\right) \nonumber\\
&&+\frac{1}{2}\left\{-2\left(1-\frac{m_c^2}{M^2}\right)\langle q^\dag i Dq\rangle_N +\frac{4m_c
}{M^2}\left(1-\frac{m_c^2}{2M^2}\right)\langle \bar{q} i D i Dq\rangle_N \right\}\exp\left(- \frac{m_c^2}{M^2}\right) \nonumber\\
&&+\frac{1}{16}\langle\frac{\alpha_sGG}{\pi}\rangle_N\int_0^1 dx \left(1+\frac{\widetilde{m}_c^2}{M^2}\right)\exp\left({-\frac{\widetilde{m}_c^2}{M^2}}\right)
-\frac{1}{48M^4}\langle\frac{\alpha_sGG}{\pi}\rangle_N\int_0^1 dx \frac{1-x}{x}\widetilde{m}_c^4  \nonumber\\
&&\exp\left({-\frac{\widetilde{m}_c^2}{M^2}}\right)\,.
\label{d0mes}
\end{eqnarray}

In above equations, $\frac{1}{M^2}$ is the Borel mass operator and  the $\langle \bar q q \rangle_N$, $\langle q^\dag i D q\rangle_N$, $\langle \bar{q} i D i Dq\rangle_{_N}$ and  $ \langle \frac{\alpha_s GG}{\pi} \rangle_N$ are the nucleon expectation values of different quark and gluon condensate. Also, $\Gamma(0,x)=e^{-x}\int_0^\infty dt \frac{1}{t+x}e^{-t}$ and $\widetilde{m}_c^2$=$m_c^2/x$. We will see later on that  as compared to scalar quark condensates $\bar{q}q$, the impact of other quark condensates (O.Q.C) $\langle q^\dag i D q\rangle_N$, $\langle \bar{q} i D i Dq\rangle_{_N}$ is very small on the observables of the $D$ mesons.

The nucleon expectation values of the chiral condensates can be calculated by using
\begin{eqnarray}
\mathcal{O}_{\rho_{N}} &=&\mathcal{O}_{vac} +
4\int\frac{d^{3}p}{(2\pi)^{3} 2 E_{p}}n_{F}\left\langle N(p)\vert \mathcal{O}\vert N(p) \right\rangle \nonumber\\
& =& \mathcal{O}_{vac} + \frac{\rho_N}{2 m_N}\mathcal{O}_{N}, 
\label{operator1}
\end{eqnarray}

where $\mathcal{O}$ denotes an operator. Now, by taking the expectation values at both sides of above equation,
\begin{eqnarray}
\langle{\cal{O}}\rangle_{\rho_N} &=&\langle{\cal{O}}\rangle_{vac}+\frac{\rho_N}{2m_N}\langle {\cal{O}}\rangle_N, \nonumber\\
 \langle{\cal{O}}\rangle_{N} &= &\frac{2m_N}{\rho_N}\left( \langle{\cal{O}}\rangle_{\rho_N}-\langle {\cal{O}}\rangle_{vac}\right) 
\end{eqnarray}
where $\langle{\cal{O}}\rangle_{vac}$ and $\langle
{\cal{O}}\rangle_N$ denote the vacuum operator and nuclear matter induced operator in the Fermi gas model,  respectively \cite{Drukarev1991}.

Following this, the nucleon expectation values of light quark and gluon condensates are expressed as, 
\begin{equation}
{<u \bar{u}>}_{N} = \left[ {<u\bar{u}>}_{\rho_{N}}  - {<u \bar{u}>}_{vac}\right] \frac{2m_N}{\rho_N},
\label{ucondexp1}
\end{equation}

\begin{equation}
{<d \bar{d}>}_{N} = \left[ {<d\bar{d}>}_{\rho_{N}}  - {<d \bar{d}>}_{vac}\right] \frac{2m_N}{\rho_N},
\label{ucondexp2}
\end{equation}

\begin{equation}
\langle \bar{q} i D i Dq\rangle_{N} = \left[ \langle \bar{q} i D i Dq\rangle_{\rho_{N}}  - \langle \bar{q} i D i Dq\rangle_{vac}\right] \frac{2m_N}{\rho_N},
\label{ucondexp3}
\end{equation}

\begin{equation}
\langle q^\dag i D q\rangle_{N} = \left[ \langle q^\dag i D q\rangle_{\rho_N}  - \langle q^\dag i D q\rangle_{vac}\right] \frac{2m_N}{\rho_N},
\label{ucondexp4}
\end{equation}

and

\begin{equation}
\left\langle  \frac{\alpha_{s}}{\pi} {G^a}_{\mu\nu} {G^a}^{\mu\nu} 
\right\rangle_{N} = \left[ \left\langle  \frac{\alpha_{s}}{\pi} {G^a}_{\mu\nu} {G^a}^{\mu\nu} \right\rangle_{\rho_N} -  \left\langle  \frac{\alpha_{s}}{\pi} {G^a}_{\mu\nu} {G^a}^{\mu\nu} \right\rangle_{vac} \right]\frac{2m_N}{\rho_N}.
\label{Gcondexp3}
\end{equation}
The condensate $\langle \bar{q} i D i Dq\rangle_{\rho_N}$ appearing in Eq.(\ref{ucondexp3}) can be calculated in terms of  light quark condensates using equations \cite{Chhabra2017,Thomas2007}

\begin{align}
\langle \bar{q} i D i Dq\rangle_{\rho_N} + \frac{1}{8}\langle\bar{q}g_s\sigma Gq\rangle_{\rho_N} =  0.3 GeV^{2}\rho_{N},
\label{cond3}
\end{align}

and

\begin{align}
\langle\bar{q}g_s\sigma Gq\rangle_{\rho_N} = \lambda^{2}\left\langle \bar{q}q \right\rangle_{\rho_{N}} + 3.0 GeV^{2}\rho_{N}.
\label{cond2}
\end{align}

In this article, we have used the condensate value $\langle q^\dag i D q\rangle_N=$0.18 GeV$^2$ $\rho_N$ from the linear density approximations results \cite{Thomas2007}. Now, in Eq.(\ref{wdiff}), to calculate the values of two unknowns $a$ and $b$, we need one more equation, which can be obtained  by  differentiation of Eq.(\ref{wdiff}) with $z=\frac{1}{M^2}$,$i.e.$,

\begin{eqnarray}
 a\, \frac{d}{dz} C_a+b\, \frac{d}{dz}C_b &=&\frac{d}{dz}C_f \,,
 \label{diff}
\end{eqnarray}
By solving  Eqs.(\ref{wdiff}) and (\ref{diff}) simultaneously, we get the following mathematical formulas to find $a$ and $b$
 \begin{eqnarray}
 a&=&\frac{C_f\left(-\frac{d}{dz}\right)C_b-C_b\left(-\frac{d}{dz}\right)C_f}{C_a\left(-\frac{d}{dz}\right)C_b-C_b\left(-\frac{d}{dz}\right)C_a}\, , \nonumber\\
  b&=&\frac{C_f\left(-\frac{d}{dz}\right)C_a-C_a\left(-\frac{d}{dz}\right)C_f}{C_b\left(-\frac{d}{dz}\right)C_a-C_a\left(-\frac{d}{dz}\right)C_b}\, .
 \end{eqnarray}
 The obtained values of $a$ and $b$  are used to calculate the mass shift  and decay constant  of $D$ mesons given by Eq.(\ref{msd}) and (\ref{fd}) respectively.

\subsection{In-medium Decay Width of Higher Charmonium States Using $^3P_0$ Model}
\label{subsec:2.3}

In the present work, one objective is to calculate the decay width of higher charmonium states ($\psi(3686),\psi(3770),{{\chi_c}_0}(3414),{{\chi_c}_2}(3556)$) to pseudo scalar $D \bar D$ mesons. In order to calculate this observable, we rely on the $^3P_0$ model\cite{Micu1969,Yaouanc1973,Yaouanc1977,Friman2002}, which is  a quark-antiquark pair creation model. In this model, light quark pair is generated in the $^3P_0$ state (vacuum), and one of the quark (antiquark) is combined with the heavy charm quark from the decaying charmonium at zero momentum. The matrix element for the decay $C$ $\rightarrow$ $D$ + $\bar D$ (where $C$ is charmonia) is given as\cite{Friman2002}
\begin{eqnarray}
M_{C \rightarrow D \bar D}&  \propto &
\int d^3k_c  \phi_C(2k_c-2k_D) \phi_D(2k_c-k_D) \phi_{\bar D}(2k_c-k_D) [\bar{u}_{k_c,s} v_{-k_c,s}].
\label{overlap}
\end{eqnarray}

In above, $k_c - k_D$ and $k_D -k_c$ represent the  momentum of charm quark and anti charm quark of charmonia $C$. Since the decaying particle is assumed to be at rest, the magnitude of the  momentum of $D$ and $\bar D$ meson is same,$i.e.$, $|k_D|=|k_{\bar D}|$.  The term $[\bar{u}_{k_c,s} v_{-k_c,s}]$ denotes the wave function of quark anti-quark pair in the vacuum and for the charmonium, We start with the harmonic oscillator wave function \cite{Friman2002,Chhabra2018}
\begin{align}
\psi_{nL{M_L}}= (-1)^n(-\iota)^L R^{L+\frac{3}{2}} \sqrt{\frac{2n!}{\Gamma(n+L+\frac{3}{2})}} \exp\Big{(}\frac{-R^2k^2}{2}\Big{)} L_n^{L+\frac{1}{2}}(R^2k^2) Y_{lm}(\bf{k}),
\label{eq:wave}
\end{align} 
 where  $L_n^{L+\frac{1}{2}}(R^2k^2)$  denotes associate Laguerre polynomial, $Y_{lm}(\bf {k})$ represents  the spherical harmonics and $R$ is the radius of the charmonia. 

Further, the decay rate of charmonium state decaying into $D \bar{D}$ pair can be represented as\cite{Chhabra2018,Kumar2010a},
\begin{eqnarray}
\Gamma (C\rightarrow D + \bar D)=
2\pi {p_D E_D E_{\bar D} \over m_C} |M_{LS}|^2\,.
\label{gammac}
\end{eqnarray}
 
 Here, $E_D$ = $\sqrt{{{m^2_D}^*} + {P^2_D}}$, $E_{\bar D}$ = $\sqrt{{{m^2_{\bar D}}^*} + {P^2_{D}}}$ and centre of mass momentum
 \begin{eqnarray}
p_D= \Bigg(\frac{ m_{C}^2}{4}- \frac{ {m_D^2}^*
 + {m_{{\bar D}}^2}^*}{2}
+\frac{( {m_D^2}^* - {m_{{\bar D}}^2}^*)^2}{4m_{C}^2}\Bigg) ^{1/2}.
\label{pd}
\end{eqnarray}
In above, $m_C$ is the mass of charmonia and $M_{LS}$  is the invariant matrix amplitude.

Using  Eq.(\ref{gammac}), the decay rate of different higher charmonium states can be represented as\cite{Friman2002,Chhabra2017,Kumar2010a}
\begin{align}
 \Gamma_{\psi(3686) \rightarrow  D{\bar D}}  =
\frac{\pi^{1/2} E_D E_{\bar {D}}}{m_\psi(3686)} \gamma ^2 \frac{2^{7} (3+2r^2)^2 (1-3r^2)^2}{3^2(1+2r^2)^7}  \nonumber\\
y^3 \Bigg(1+ \frac{2r^2 (1+r^2)}{(1+2r^2)(3+2r^2)(1-3r^2)}y^2\Bigg)^2  e^{-\frac{y^2}{2(1+2r^2)}},
\label{eqdw1}
\end{align}
\begin{align}
 \Gamma_{\psi(3770) \rightarrow  D{\bar D}}  =
 \frac{\pi^{1/2} E_D E_{\bar {D}}}{m_\psi(3770)} \gamma ^2 \frac{2^{11} 5}{3^2} \Bigg(\frac{r}{1+2r^2}\Bigg)^7 y^3 
\Bigg(1- \frac{1+r^2}{5(1+2r^2)} y^2\Bigg)^2 e^{-\frac{y^2}{2(1+2r^2)}},
\label{eqdw2}
\end{align}

 \begin{eqnarray}
  \Gamma_{\chi_{c0}(3414) \rightarrow  D{\bar D}} 
&=& \pi^{1/2} \frac{E_D E_{\bar {D}}}{2 m_{\chi_{c0}}(3414)}\gamma^2 2^{9} 
3\Bigg(\frac{r}{1+2r^2}\Bigg)^5 
y\Bigg(1-\frac{1+r^2}{3(1+2r^2)}y^2\Bigg)^2 
 e^{-\frac{y^2}{2(1+2r^2)}},
 \label{eqdw3}
 \end{eqnarray}

and

\begin{align}
  \Gamma_{\chi_{c2}(3556) \rightarrow  D{\bar D}}  &= \frac{\pi^{1/2} E_D E_{\bar {D}}}{m_{\chi_{c2}}(3556)} \gamma ^2 \frac{2^{10} r^5 (1+r^2)^2}{15(1+2r^2)^7} y^5 e^{-\frac{y^2}{2(1+2r^2)}}.
\label{eqdw4}
\end{align}
 
In above equations, the variables $r$ and $\beta_D$ incorporate the modification of wave function due to the nodal structure of the initial and final state mesons \cite{Eichten1978,Vijande2005} and their values are fitted with the help of the experimental partial decay width of $\psi(4040)$ to $D \bar D$ mesons \cite{Friman2002}. The parameter $y$ is  expressed as,
 $y$ = $\frac{p_D}{\beta_D}$ and the  parameter $\gamma$ denotes  the strength of the vertex and fitted using the  experimental decay width of $\Gamma$($\psi(3770)$ $\rightarrow$ $D \bar{D}$)\cite{Chhabra2017,Friman2002}. The decay width of different higher charmonia can be calculated by the above equations by using the effective mass of $D$ meson obtained from the QCDSR calculations.

\section{Numerical Results and Discussions}
\label{sec:3}

 \begin{table}
\begin{tabular}{|c|c|c|c|c|}

\hline
$k_0$ & $k_1$ & $k_2$ & $k_3$ & $k_4$  \\ 
\hline 
2.53 & 1.35 & -4.77 & -2.77 & -0.218  \\ 
\hline

\hline 
$\sigma_0$ (MeV)& $\zeta_0$ (MeV) & $\chi_0$ (MeV)  & $d$ & $\rho_0$ ($\text{fm}^{-3}$)  \\ 
\hline 
-93.29 & -106.8 & 409.8 & 0.064 & 0.15  \\ 
\hline

\hline 
$m_\pi $ (MeV) &$ m_K$ (MeV)&$ f_\pi$ (MeV)  & $f_K$ (MeV) & $g_4$ \\ 
\hline 
139 & 498 & 93.29 & 122.14 & 79.91  \\ 
\hline

\hline
$g_{\sigma N}$  & $g_{\zeta N }$  &  $g_{\delta N }$  &
$g_{\omega N}$ & $g_{\rho N}$ \\

\hline 
10.56 & -0.46 & 2.48 & 13.35 & 5.48  \\ 
\hline

\end{tabular}
\caption{Values of different parameters.} \label{ccc}
\end{table} 
 
%
%
%
%
%

 In this section, we will discuss our observations on effect of magnetic field on masses and decay constant   of  pseudoscalar ($D^+$,$D^0$) and scalar ($D^+_0$,$D^0_0$) mesons and in-medium decay width of higher charmonium states ($\psi(3686),\psi(3770),{{\chi_c}_0}(3414)$ and ${{\chi_c}_2}(3556)$) in asymmetric nuclear matter at finite temperature. As discussed earlier, the light quark condensates and gluon condensates have been calculated by using chiral SU(3) model and the different parameters used in model are tabulated in Table \ref{ccc}. In addition to these, the value of charm quark mass $m_c$, running coupling constant $\alpha_s$, coupling constant $g_{DNH}$ and constant $\lambda$ are approximated to be  1.3 GeV, 0.45, 6.74 and 0.5, respectively \cite{Chhabra2017,Wang2015}. The vacuum masses of $D$ mesons are taken as 1.869, 1.864, 2.355 and 2.350 GeV for $D^+$, $D^0$, ${D^+_0}$ and  ${D^0_0}$ mesons,  respectively. The vacuum values of the decay constant for pseudoscalar and scalar mesons are taken as 0.210 and 0.334 GeV, respectively \cite{Chhabra2017,Chhabra2018}. Furthermore, the continuum threshold parameter $s_0$ for pseudoscalar and scalar mesons are taken as 6.2 and 8 GeV$^2$, respectively \cite{Wang2015}. We have chosen the proper Borel window so that there will be a least variation in the mass shift ($\Delta m^*_D$) and decay shift ($\Delta f^*_D$). The Borel Window for masses of ($D^+$,$D^0$) and  ($D^+_0$,$D^0_0$) are taken as  (4.5-5.5)  and(6-7) GeV$^2$, respectively, whereas the range of Borel window for decay constant of ($D^+$,$ D^0$) and  ($D^+_0$,$D^0_0$) are taken as (2-3)    and (7-9) GeV$^2$, respectively. Moreover, the values of parameters $\gamma$, $\beta_D$ and $r$ used in $^3P_0$ model are taken as 0.281, 0.30 and 1.04, respectively \cite{Chhabra2017}. Our further discussion of this section is divided into three subsections.
 
 \subsection{Magnetic Field Induced Quark and Gluon Condensates.}
 \label{subsec:3.1}

\begin{figure}
\includegraphics[width=16cm,height=21cm]{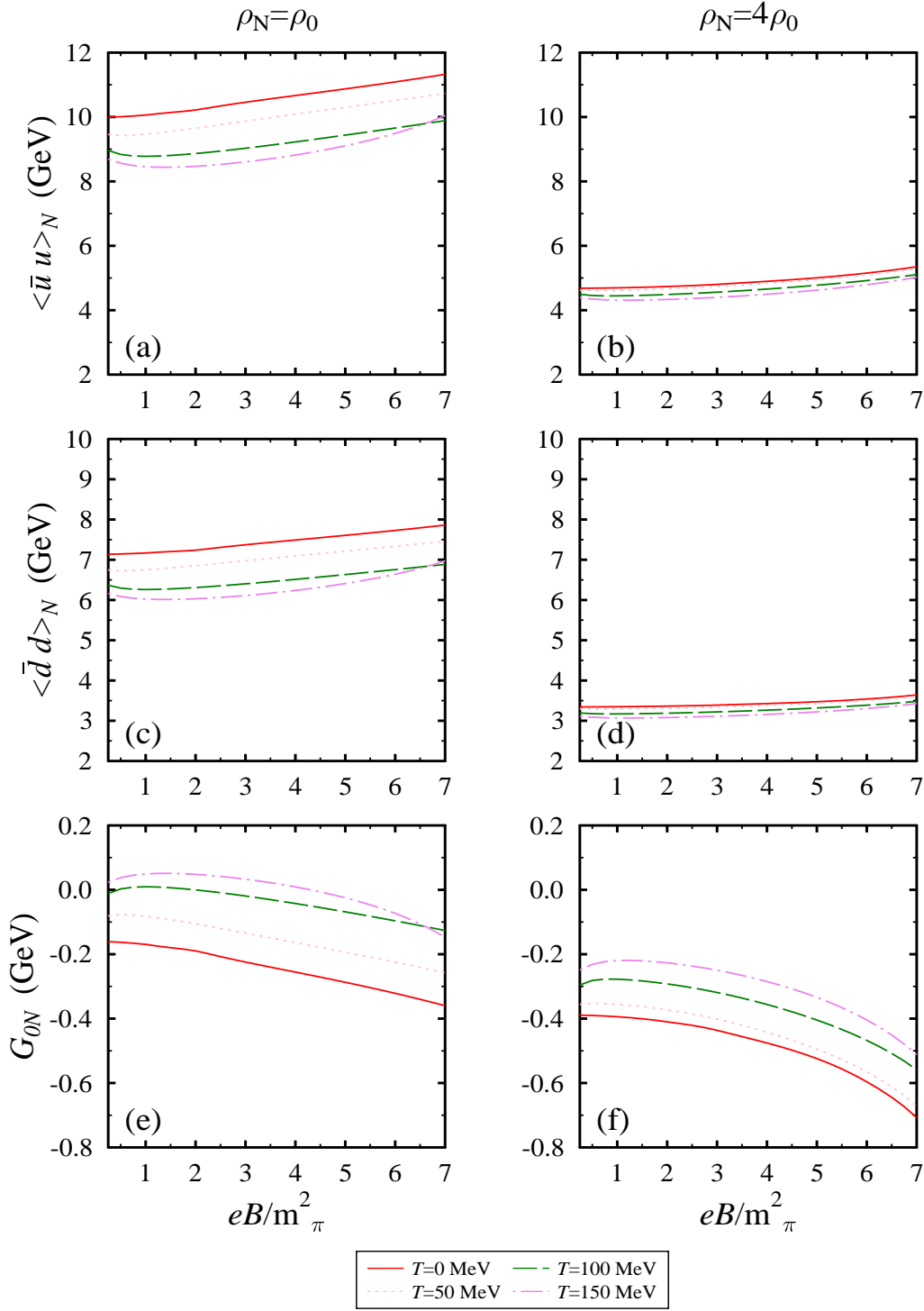}
\caption{(Color online)The nucleon expectation value of light quark condensates ($\langle u \bar u \rangle_N$ and $\langle d \bar d \rangle_N$) and gluon condensate $\left\langle  \frac{\alpha_{s}}{\pi} {G^a}_{\mu\nu} {G^a}^{\mu\nu} 
\right\rangle_{N}$  (denoted by $G_{0N}$) is plotted for symmetric nuclear matter ($\eta=0$) as a function of magnetic field $eB$ under different conditions of medium. }
\label{ce0}
\end{figure}

\begin{figure}
\includegraphics[width=16cm,height=21cm]{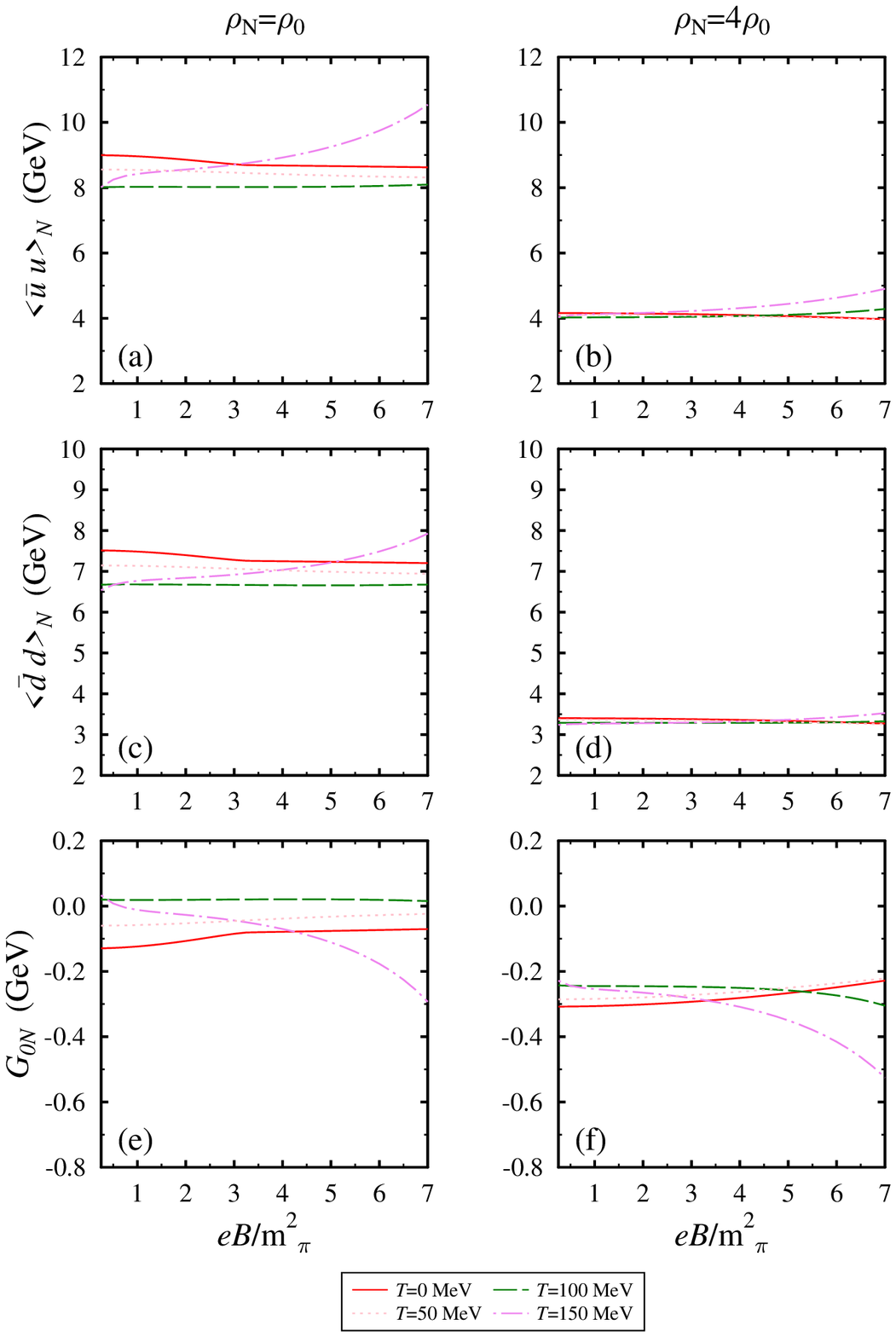}
\caption{(Color online) The nucleon expectation value of light quark condensates ($\langle u \bar u \rangle_N$ and $\langle d \bar d \rangle_N$) and gluon condensate $\left\langle  \frac{\alpha_{s}}{\pi} {G^a}_{\mu\nu} {G^a}^{\mu\nu} 
\right\rangle_{N}$  (denoted by $G_{0N}$) is plotted for asymmetric nuclear matter ($\eta \neq 0$) as a function of magnetic field $eB$ under different conditions of medium. }
\label{ce5}
\end{figure}

 In this subsection, we have shown the results for the medium induced light quark and gluon condensates, which are calculated using chiral model described in subsect \ref{subsec:2.1}. From the expressions given in Eqs.(\ref{qu})-(\ref{chiglum}), one can see that the condensates depend upon the scalar fields $\sigma$, $\zeta$, $\delta$ and $\chi$, which are solved under different conditions of medium such as density, magnetic field, temperature and asymmetry \cite{Kumar2019,Kumar2019a}. In \cref{ce0} and \cref{ce5}, we have plotted the nucleon expectation values of scalar up quark condensate $\langle u \bar u \rangle_N$, down quark condensate  $\langle d \bar d \rangle_N$ and gluon condensate ($\left\langle  \frac{\alpha_{s}}{\pi} {G^a}_{\mu\nu} {G^a}^{\mu\nu} 
\right\rangle_{N}$)  with respect to magnetic field at isospin symmetry parameters $\eta$=0 and  0.5, respectively. We have shown the results at nucleon densities $\rho_N$=$\rho_0$ and $4\rho_0$, and temperatures $T$ = 0, 50, 100 and 150 MeV. In \cref{ce0}, at $\eta=0$, we can see that the magnitude of up and down quark condensates increases with the increase in the magnetic field. One can also conclude that the density effects are also appreciable as the magnitude of quark condensates decrease with the increase in the density. Moreover, inclusion of temperature effects decrease the magnitude of quark condensates but the trend concerning magnetic field remains the same. For example, in symmetric nuclear medium for $eB=4m^2_\pi$, the values of $\langle u \bar u \rangle_N$ ($\langle d \bar d \rangle_N$), at $\rho_N=\rho_0$, is 10.67 (7.46), 9.22 (6.48), 8.81 (6.20) GeV for $T$=0,100 and 150 MeV, respectively and for  $\rho_N=4\rho_0$ it modifies to 4.9 (3.41), 4.7 (3.25) and 4.5 (3.14) GeV. It may be noted that despite $\eta$=0, the value of up and down quark condensates are different which is contradictory to the previous work (at zero magnetic field) as up and down quark are isospin partner and hence indistinguishable in symmetric nuclear matter \cite{Kumar2015}. This is because of the Landau quantization, which occurs due to the interaction of charged protons with magnetic field. This interaction disturbs the equality between scalar density of proton and neutron and hence the magnitude of $\delta$ field become non-zero \cite{Kumar2019a}. 

In sub-plots (e) and (f), we have also shown the results for scalar gluon condensate and observed that the magnitude of gluon condensate changes very less as compared to light quark condensates. This is the reason why open charm meson experience larger mass shift than the ground state  charmonia \cite{Kumar2019,Chhabra2017}. The value of gluon condensate at nuclear saturation density decrease as a function of magnetic field and this trend becomes more appreciable at $\rho_N$=$4\rho_0$. Furthermore, we observe the temperature effects on gluon condensate are  opposite than the quark condensate. This is because the gluon condensate depends on the fourth power of $\chi$ field  along with $\sigma$ and $\zeta$ field (see Eq.\ref{chiglum}) whereas quark condensates (see Eq.\ref{qu} and \ref{qd}) has only $\sigma$ and $\delta$ field dependence \cite{Kumar2019a}. 

In \cref{ce5}, at $\eta$=0.5, we observe similar behaviour of condensates as a function of magnetic field as was at $\eta = 0$, expect at low temperature.  For example, at $eB=4m^2_\pi$, the values of $\langle u \bar u \rangle_N$ ($\langle d \bar d \rangle_N$) at $\rho_N=\rho_0$ is 8.67 (7.25), 8.02 (6.66), 8.92 (7) GeV for $T$=0,100 and 150 MeV, respectively and for  $\rho_N=4\rho_0$ it modifies to 4.1 (3.36), 4.06 (3.28) and 4.31 (3.31) GeV. We also observed a crossover behaviour in the plot of gluon condensate. In this case, the  expectation value of scalar gluon condensate  increase for low temperature whereas it decrease for high temperature. This is because in a medium having large number of neutrons, the behaviour of neutron scalar density in low-temperature modifies \cite{Kumar2019a,Reddy2018}.  The nucleon expectation values of condensates are linked with nuclear matter  expectation value of condensates via Eqs.(\ref{ucondexp1}), (\ref{ucondexp2}) and (\ref{Gcondexp3}) Therefore, from Eqs.(\ref{qu}) and (\ref{qd}), one can see that in symmetric nuclear matter, these quark condensates are directly proportional to scalar fields $\sigma$ and $\delta$, hence the behaviour of $q \bar q$ with different medium parameters is same as that of $\sigma$ field (as $\delta$ field has very less variation with magnetic field for $\eta = 0$) \cite{Kumar2019a}. On the other hand, for asymmetric matter, the condensates have mixed contributions of $\sigma$ and $\delta$ fields as the $\delta$ field varies appreciably with the increase in magnetic field.  Moreover, the gluon condensate has the dependence on scalar fields $\sigma$, $\zeta$ and $\chi$. The scalar fields as a function of magnetic field with different value of density, asymmetry and temperature are plotted and discussed in our previous work \cite{Kumar2019a}. As per inverse magnetic catalysis, the scalar fields gets enhanced due to the generation of additional fermion anti-fermion condensates in the presence of magnetic field \cite{Kharzeev2013,Bali2012,Kumar2019a}.


  \subsection{Mass and Shift in Decay Constant of Pseudoscalar and Scalar $D$ Mesons.}
 \label{subsec:3.2}

\begin{figure}
\includegraphics[width=16cm,height=21cm]{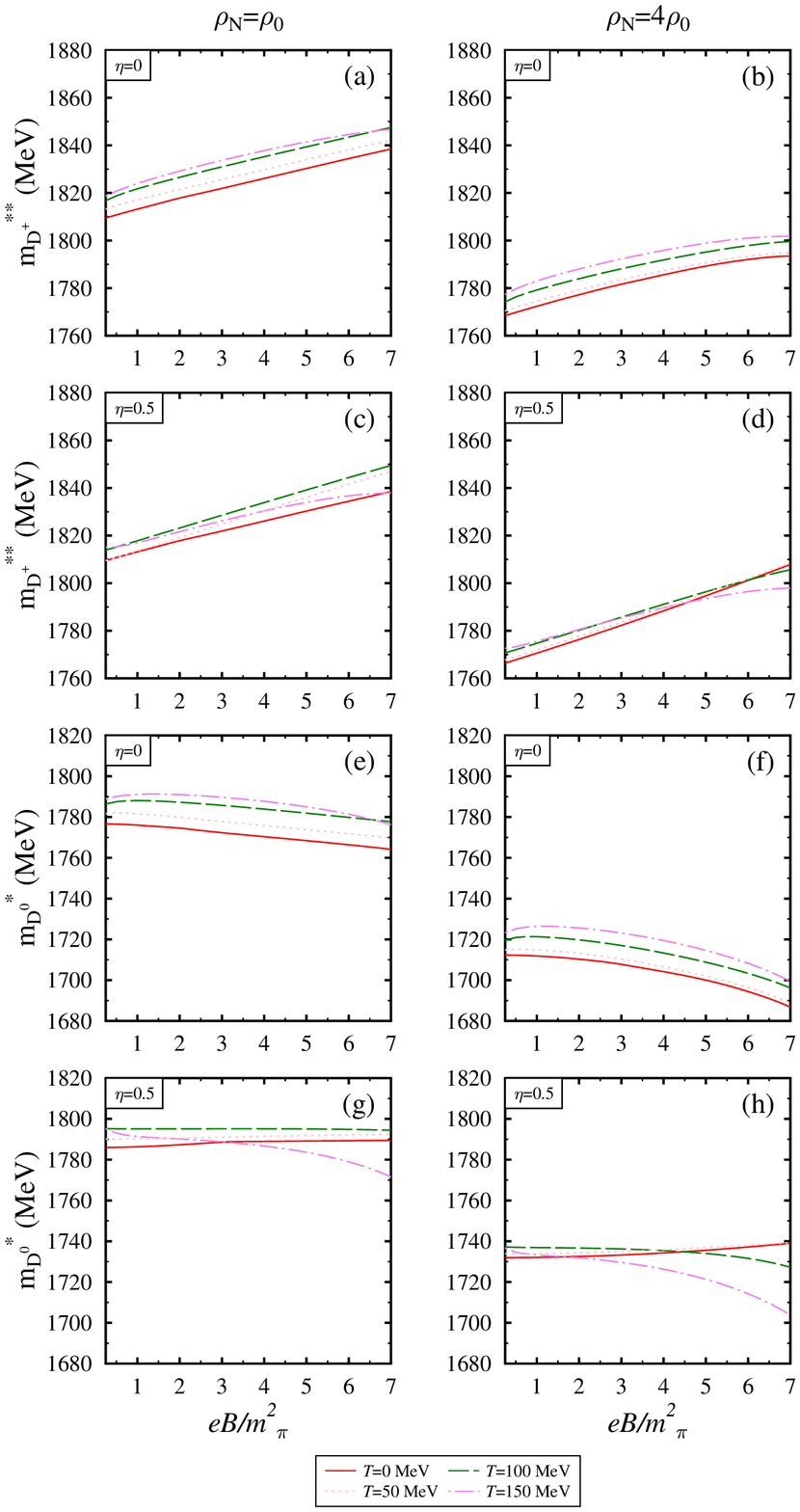}
\caption{(Color online) The effective mass of pseudoscalar $D^+$ (charged) and  $D^0$ (uncharged)  mesons is plotted as a function of magnetic field $eB$ under different conditions of medium. }
\label{ms_ps}
\end{figure}

\begin{figure}
\includegraphics[width=16cm,height=21cm]{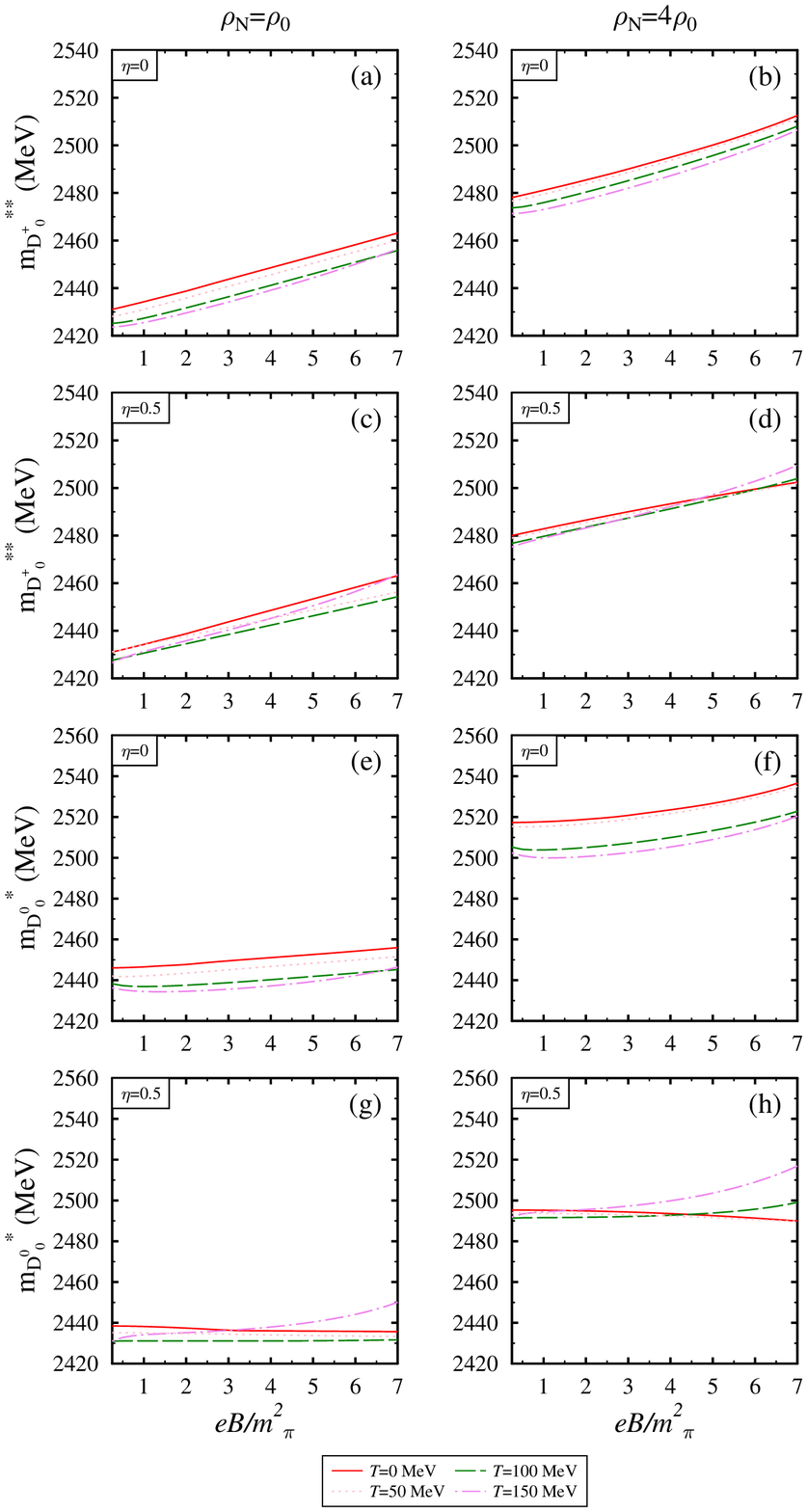}
\caption{(Color online) The effective mass of scalar $D^+_0$ (charged) and  $D^0_0$ (uncharged)  mesons is plotted as a function of magnetic field $eB$ under different conditions of medium. }
\label{ms_s}
\end{figure}

 Here we will discuss how mass and shift in decay constant of isospin averaged pseudoscalar ($D^+,D^0$) and scalar ($D^+_0$,$D^0_0$) mesons modifies with magnetic field and other medium properties. As discussed earlier in subsect \ref{subsec:2.2}, the above observables are calculated in the QCDSR by using quark and gluon condensates. In \cref{ms_ps} and \cref{ms_s}, we have plotted the masses of pseudoscalar and scalar $D$ mesons, respectively, as a function of external magnetic field in  nuclear matter. The values of the effective mass of pseudoscalar and scalar $D$ mesons in the presence of magnetic field and other medium properties are shown in Table \ref{tablems} for better comparison. In \cref{ms_ps}, for symmetric nuclear matter, we observe that the effective mass of charged  $D^+$ meson increase with the increase in magnetic field and hence, the magnitude of  negative mass shift of $D^+$ meson decreases. For example, at $\eta=0$, $\rho_N=\rho_0$ and $T=0$, the value of mass increase from   1826 to 1834 MeV,  when we move from $eB=4m^2_\pi$ to $6m^2_\pi$. This explanation lies to the fact that the $D^+$ meson is a charged meson and with the interaction of magnetic field additional positive mass shift (see Eq.(\ref{mdpm_landau})) comes into picture. If we do not consider the additional mass shift, then we observe a negative  in-medium mass shift which increases with the increase in magnetic field. The density effect on the mass shift of $D^+$ meson is also very prominent as the mass shift of $D^+$ meson increase with the increase in density. For instance, at $eB=4m^2_\pi$, $\eta=0$ and $T$=0 MeV, when we move from $\rho_N=\rho_0$ to  $4\rho_0$, the values of mass changes from 1826 to 1786 MeV.  The inclusion of temperature effects increases the mass at particular value of density.
 For example, at $\eta=0$, $eB=4m^2_\pi$,  and $\rho_N=4\rho_0$, when we move from $T$=0 to  100 MeV, the values of mass changes from 1786 to 1792 MeV. 
 
 For  neutral pseudoscalar $D^0$ meson there will not be any  additional positive mass shift (see Eq.(\ref{mdpm_landau})) and therefore the effective mass in this case does not increase but decrease with the increase in magnetic field (see \cref{tablems}).  However, the temperature and density effects remain same as for $D^+$ meson. For instance, at $eB=4m^2_\pi$, $\eta=0$ and $T$=0 (100) MeV, the mass of $D^0$ meson is 1770 (1784) and 1704 (1713) MeV for  $\rho_N=\rho_0$ and  $4\rho_0$ respectively and for  $eB=6m^2_\pi$, the value  changes to 1766 (1780)  and 1694 (1703) MeV.  It is observed that for both pseudoscalar mesons, if we go from symmetric matter to highly asymmetric matter, the effect of temperature become less appreciable except for $T$=150 MeV.  The isospin asymmetry effects should be quite visible for $D^+$ ($c \bar d$) and $D^0$ ($c \bar u$) mesons as they are isospin partner of each other. For non-magnetic  nuclear matter, appreciable isospin effects are observed \cite{Chhabra2017}. In \cite{Chhabra2017}, the mass of $D^0$ meson found to be decrease whereas mass of $D^+$ increases as we go from symmetric to asymmetric nuclear matter.  But in the present case, the asymmetry effects on $D^+$ meson are compensated  by the additional positive mass shift by Landau interaction hence we see less crossover temperature effects. On the other hand,  the effective mass of $D^0$ meson modifies appreciably in asymmetric nuclear matter and show crossover behaviour. It shows slight increase for high temperature but for  low temperature it decrease appreciably. For example, at $eB=6m^2_\pi$, $\eta=0.5$ and $T$=0 (100) MeV, the mass of $D^0$ meson is 1789 (1794) and  1737 (1731) MeV for  $\rho_N=\rho_0$ and  $4\rho_0$ respectively. This crossover is a reflection of behaviour of quark and gluon condensates in asymmetric magnetized nuclear matter as shown in \cref{ce5}.
  
As can be seen from  \cref{ms_s}, contrary to pseudoscalar $D$ mesons, we observed positive mass shift for  scalar
$D^+_0$ and  $D^0_0$  mesons. The fact that the  effective mass will decrease or increase  depends upon the sign of scattering length (see Eq.(\ref{slength})). In heavy meson-nucleon bound states, the  negative or positive sign of scattering length determines  whether the $DN$ interactions are attractive or repulsive \cite{Wang2015}. The enhancement in mass of $D^+_0$ meson is more than the $D^0_0$ meson as a function of magnetic field. As discussed earlier, the charged meson experiences additional positive shift due to induced Landau levels. For scalar $D$ mesons,  we observed an opposite behaviour of  masses as a function of temperature comparative to  pseudoscalar case.  In symmetric nuclear matter,  the effective mass of scalar $D$ meson decrease with the increase in temperature while follows the same trend as a function of magnetic field. The asymmetric effects on scalar $D^+_0$ meson are also compensated by the magnetic field as was for  pseudoscalar case. 

As we pointed earlier also, condensates $\langle q^\dag i Dq \rangle_N$ and  $\langle \bar{q} i D i Dq\rangle_N$ have less effect as compared to $\bar{q}q$ condensates on in-medium properties of charmed mesons (see \cref{tablepsc} and \ref{tablesc}
for pseudo scalar and scalar mesons respectively). In these tables, we have also compared the mass shift with and without magnetic effect. The results of mass shift in the absence magnetic field have been taken from the Ref. \cite{Chhabra2017} (pseudoscalar) and \cite{Chhabra2018} (scalar). In this comparison, we observed that the presence of magnetic field affects the mass shift of charged pseudoscalar and scalar $D$ meson significantly, whereas for neutral $D$ meson the effects are good but not prominent as compared to charged meson. This is due to additional positive mass shift in the medium which occur via Landau quantization as discussed earlier also.

In our knowledge, the in-medium masses of $D$  mesons at finite temperature and density of nuclear matter considering external magnetic fields have not been evaluated yet within any model.
 In \cite{Reddy2018} the mass of pseudoscalar $D$ meson in strongly magnetized asymmetric cold nuclear matter has been calculated solely in the effective chiral SU(4) model. In this article, the mass splitting between $D^+$ ($D^0$) and  $D^-$ ($ \bar D^0$) has been calculated using the self-energy of $D$ mesons. In cold and asymmetric nuclear matter, the mass shift of $D$ mesons has been compared with and without taking the contributions from Anomalous Magnetic Moment (AMM).
  In Ref. \cite{Gubler2016}, authors used hadronic QCD sum rules in which the magnetic field effects are introduced  on both phenomenological as well as OPE side to evalauae the effect of magnetic field on $D$ mesons properties (at zero density and temperature). The additional mixing effects are examined on the phenomenological side by adding spectral \textit{ansatz} term.  Besides the mixing effects, an additional perturbative positive mass shift is also found due to the magnetic field. In our future work, we will also include the mixing effects of $D$ mesons in the presence of magnetic field.
 
\begin{figure}
\includegraphics[width=16cm,height=21cm]{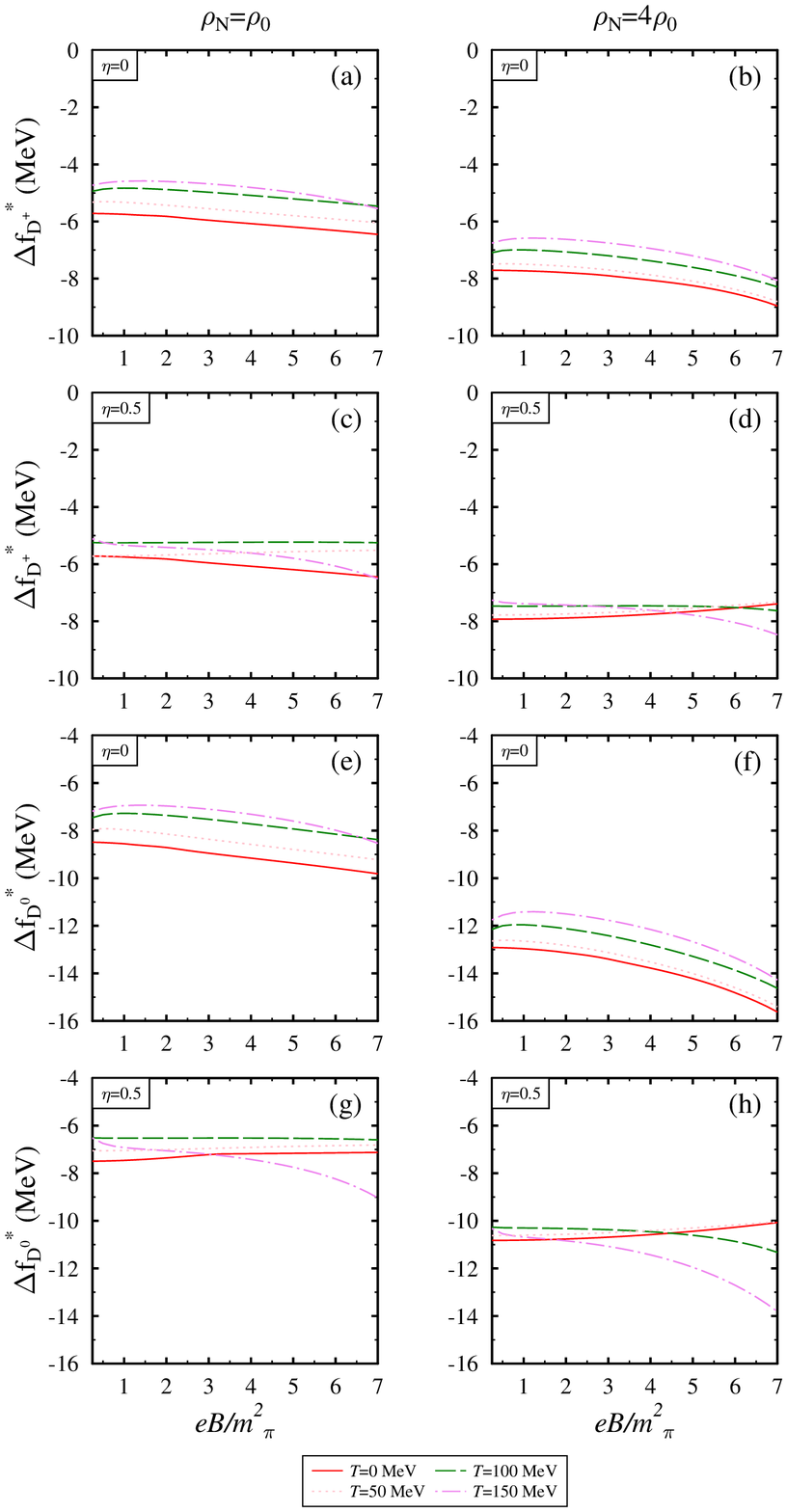}
\caption{(Color online) The shift in decay constant of pseudoscalar $D^+$ (charged) and  $D^0$ (uncharged)   mesons is plotted as a function of magnetic field $eB$ under different conditions of medium. }
\label{fd_ps}
\end{figure}

\begin{figure}
\includegraphics[width=16cm,height=21cm]{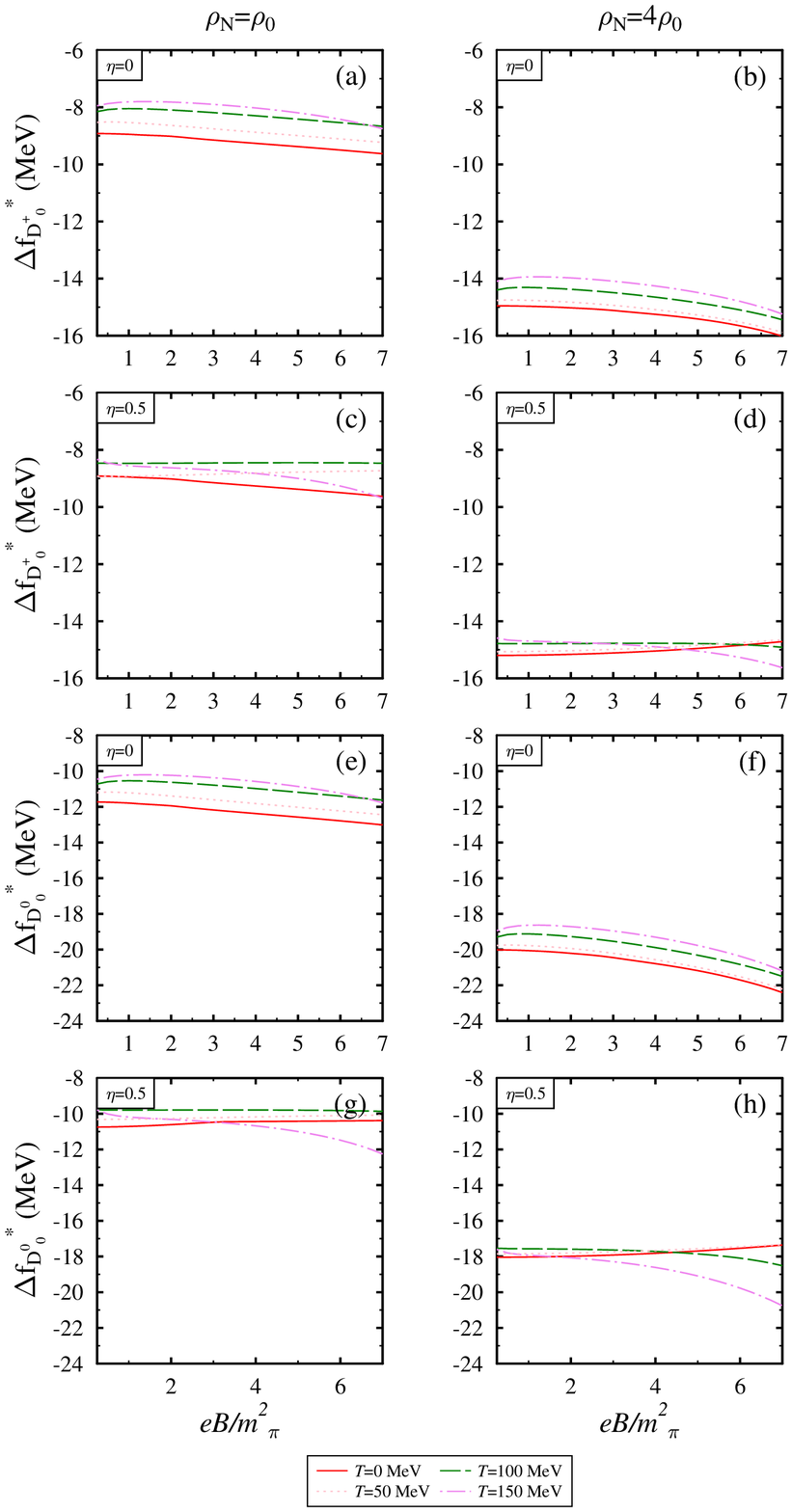}
\caption{(Color online) The shift in decay constant of scalar $D^+_0$ (charged) and  $D^0_0$ (uncharged)   mesons is plotted as a function of magnetic field $eB$ under different conditions of medium. }
\label{fd_s}
\end{figure}

 In \cref{fd_ps} (\cref{fd_s}), for a given value of density, temperature and isospin asymmetry the shift in decay constant $\Delta f_D^*$ of pseudoscalar (scalar) $D$ meson is plotted as a function of magnetic field. The values of in-medium shift in decay constant of pseudoscalar and scalar $D$ meson in the presence of magnetic field and other medium properties are shown in Table \ref{tablefd}. In symmetric nuclear matter,  the magnitude of $\Delta f_D^*$ increase as a function of magnetic field, whereas it decrease with the increase in temperature. But for $\eta=0.5$, the crossover behaviour is observed as was the case for effective masses. This is because the shift in decay constant is calculated using the shift in effective mass (see Eq.(\ref{fd})) in the QCDSR. It may be noted that for scalar mesons, the shift in decay constant is negative despite the positive mass shift.

 \begin{table}
\begin{tabular}{|c|c|c|c|c|c|c|c|c|c|}
\hline
& & \multicolumn{4}{c|}{$\eta$=0}    & \multicolumn{4}{c|}{$\eta$=0.5}   \\
\cline{3-10}
&$eB/{{m}_{\pi}^2}$ & \multicolumn{2}{c|}{T=0} & \multicolumn{2}{c|}{T=100 }& \multicolumn{2}{c|}{T=0}& \multicolumn{2}{c|}{T=100 }\\
\cline{3-10}
&  &$\rho_0$&$4\rho_0$ &$\rho_0$  &$4\rho_0$ & $\rho_0$ &$4\rho_0$&$\rho_0$&$4\rho_0$ \\ \hline 
$ m^{**}_{D^+}$& 4&1826& 1786& 1835 &1792&1828&1788&1833&1791 \\ \cline{2-10}
&6&1834&1792  &1843  & 1798 & 1839 &1801 &1844 & 1801 \\ \cline{1-10}
$m^*_{D^0}$&4&1770 & 1704 & 1784 &1713 &1779 &1734 & 1795 &1735 \\  \cline{2-10}
&6&1766 &1694 &1780 &1703 & 1789 &1737 & 1794&1731 \\  
 \cline{1-10}
$ m^{**}_{D_0^+}$&4&2448 & 2495 & 2441 &2490 &2446 &2493 & 2442 &2491 \\  \cline{2-10}
&6&2458 &2505 &2450 &2501 & 2454 &2494 &2450&2499\\  \hline
$ m^*_{D_0^0}$&4&2451 & 2516 & 2440 &2509 &2436 &2493 & 2431 &2492 \\  \cline{2-10}
&6&2459 &2524 &2443 &2517 & 2435 &2491 &2431&2495\\  \cline{1-10}
\end{tabular}
\caption{In above, we tabulate the values of magnetic field induced masses  of $D^+$, $D^0$, $D_0^0$ and $D_0^+$ mesons (in units of MeV).}
\label{tablems}
\end{table}

 \begin{table}
\begin{tabular}{|c|c|c|c|c|c|c|c|c|c|c|c|}
\hline
& & & \multicolumn{4}{c|}{$\eta$=0} & \multicolumn{4} {c|}{$\eta$=0.5}\\  \cline{4-11}

& & & \multicolumn{2}{c|}{T=0} &\multicolumn{2}{c|}{T=100 }& \multicolumn{2}{c|}{T=0} & \multicolumn{2}{c|}{T=100 }    \\\cline{4-11}

& & & $\rho_0$ & $4\rho_0$  &$\rho_0$&4$\rho_0$&$\rho_0$&4$\rho_0$ & $\rho_0$ & 4$\rho_0$ \\ \hline
&All Condensates  &$\Delta m^{**}$&-43&-83&-34 &-77 &-41 &-81 & -36 & -78\\ \cline{3-11}
$D^+$&  & $\Delta m (B=0)$ &-64&-110&-55 &-103 &-68 &-112 & -60 & -108 \\ \cline{2-11}

&O.Q.C=0 &$\Delta m^{**}$&-60&-92& -51 &-85 &58 & -89 & -52 &-86\\ \cline{3-11}
 && $\Delta m (B=0)$ &-62 &-101 & -53 & -94 &-66  &-104  &-58 &-99\\ \hline
 
 &All Condensates  &$\Delta m^{*}$&-94&-160&-80 &-151 &-85 &-130 & -69 & -129\\ \cline{3-11}
$D^0$&  & $\Delta m (B=0)$ &-92&-163&-79&-153 &-81 &-141 & -72 & -137 \\ \cline{2-11}

&O.Q.C=0 &$\Delta m^{*}$&-89&-146&-76 &-137 &-71 & -116 & -65 &-115\\ \cline{3-11}
 && $\Delta m (B=0)$ &-90 &-154 & -77 & -144 &-79  &-133  &-69 &-128\\ \hline

\end{tabular}
\caption{In the above table magnetic field induced mass shift of $D^{+}$ and $D^0$ mesons (in MeV) at $eB=4{{m}_{\pi}^2}$ are compared with the mass shift obtained without magnetic field \cite{Chhabra2017}.  We have also considered and compared the contribution of other quark condensates (O.Q.C).}
\label{tablepsc}
\end{table}

  \begin{table}
\begin{tabular}{|c|c|c|c|c|c|c|c|c|c|c|c|}
\hline
& & & \multicolumn{4}{c|}{$\eta$=0} & \multicolumn{4} {c|}{$\eta$=0.5}\\  \cline{4-11}

& & & \multicolumn{2}{c|}{T=0} &\multicolumn{2}{c|}{T=100 }& \multicolumn{2}{c|}{T=0} & \multicolumn{2}{c|}{T=100 }    \\\cline{4-11}

& & & $\rho_0$ & $4\rho_0$  &$\rho_0$&4$\rho_0$&$\rho_0$&4$\rho_0$ & $\rho_0$ & 4$\rho_0$ \\ \hline
&All Condensates  &$\Delta m^{**}$&93&140&86 &135&91 &114 & 87 & 136\\ \cline{3-11}
$D^+_0$&  & $\Delta m (B=0)$ &64&125&58 &120 &68 &127 & 62 & 123 \\ \cline{2-11}

&O.Q.C=0  &$\Delta m^{**}$&87&133&80 &128 &85 & 131 & 81 &129\\ \cline{3-11}
 && $\Delta m (B=0)$ &66 &138 & 59 & 132 &70  &140  &63 &137\\ \hline
 
 &All Condensates  &$\Delta m^{*}$&101&166&90 &159 &86 &143 & 81 & 142\\ \cline{3-11}
$D^0_0$&  & $\Delta m (B=0)$ &87&162&76 &156 &78 &148 & 72 & 143 \\ \cline{2-11}

&O.Q.C=0  &$\Delta m^{*}$&110&174&99 &168 &95 & 156 & 90 &151\\ \cline{3-11}
 && $\Delta m (B=0)$ &89&180 & 79 & 173 &81  &164 &73 &160\\ \hline

\end{tabular}
\caption{In the above table magnetic field induced mass shift of $D^{+}_0$ and $D^0_0$ mesons (in MeV) at $eB=4{{m}_{\pi}^2}$ are compared with the mass shift obtained without magnetic field \cite{Chhabra2018}.  We have also considered and compared the contribution of additional condensates.}
\label{tablesc}
\end{table}

 \begin{table}
\begin{tabular}{|c|c|c|c|c|c|c|c|c|c|}
\hline
& & \multicolumn{4}{c|}{$\eta$=0}    & \multicolumn{4}{c|}{$\eta$=0.5}   \\
\cline{3-10}
&$eB/{{m}_{\pi}^2}$ & \multicolumn{2}{c|}{T=0} & \multicolumn{2}{c|}{T=100 }& \multicolumn{2}{c|}{T=0}& \multicolumn{2}{c|}{T=100 }\\
\cline{3-10}
&  &$\rho_0$&$4\rho_0$ &$\rho_0$  &$4\rho_0$ & $\rho_0$ &$4\rho_0$&$\rho_0$&$4\rho_0$ \\ \hline 
$ \Delta f^{*}_{D^+}$& 4& 6.07&-8.05& -5.18 &-7.38&-5.85&-8.05&-5.08&-7.38\\ \cline{2-10}
&6&-6.31  &-8.53  & -5.43 & -7.89 &5.79 &-8.53& -5.33 & -7.89 \\ \cline{1-10}
$ \Delta f^*_{D_0^0}$&4&-9.15 & -13.27 & -7.71 &-12.80 &-7.17 &-10.57 & -6.52 &-10.45 \\  \cline{2-10}
&6&-9.51 &-14.81 &-8.14 &-13.86 & -7.14 &-10.26 & -6.55&-10.86 \\  
 \cline{1-10}
$  \Delta f^{*}_{D_0^+}$&4&-9.26 & -15.25 & -8.29 &-14.65 &-9.03 &-15.04&-8.45 & -14.77  \\  \cline{2-10}
&6&-9.49 &-15.65 &-8.53 &-15.09 & -9.01 &-14.84 &-8.45&-14.81 \\  \hline
$  \Delta f^*_{D_0^0}$& 4&-12.37&-20.74 &-10.97&-19.88&-10.43&-17.81&-9.73 &-17.71\\ \cline{2-10}
&6&-12.78  &-21.66  &-11.39 & -20.83&-10.40 &-17.53 & -9.82 & -18.09 \\ \cline{1-10}
\end{tabular}
\caption{In above, we tabulate the values of magnetic field induced shift in decay constants of $D^+$, $D^0$, $D_0^0$ and $D_0^+$ mesons (in units of MeV).}
\label{tablefd}
\end{table}
 
 \subsection{In-medium Decay Width of Higher Charmonium States}
 \label{subsec:3.3}

\begin{figure}
\includegraphics[width=16cm,height=21cm]{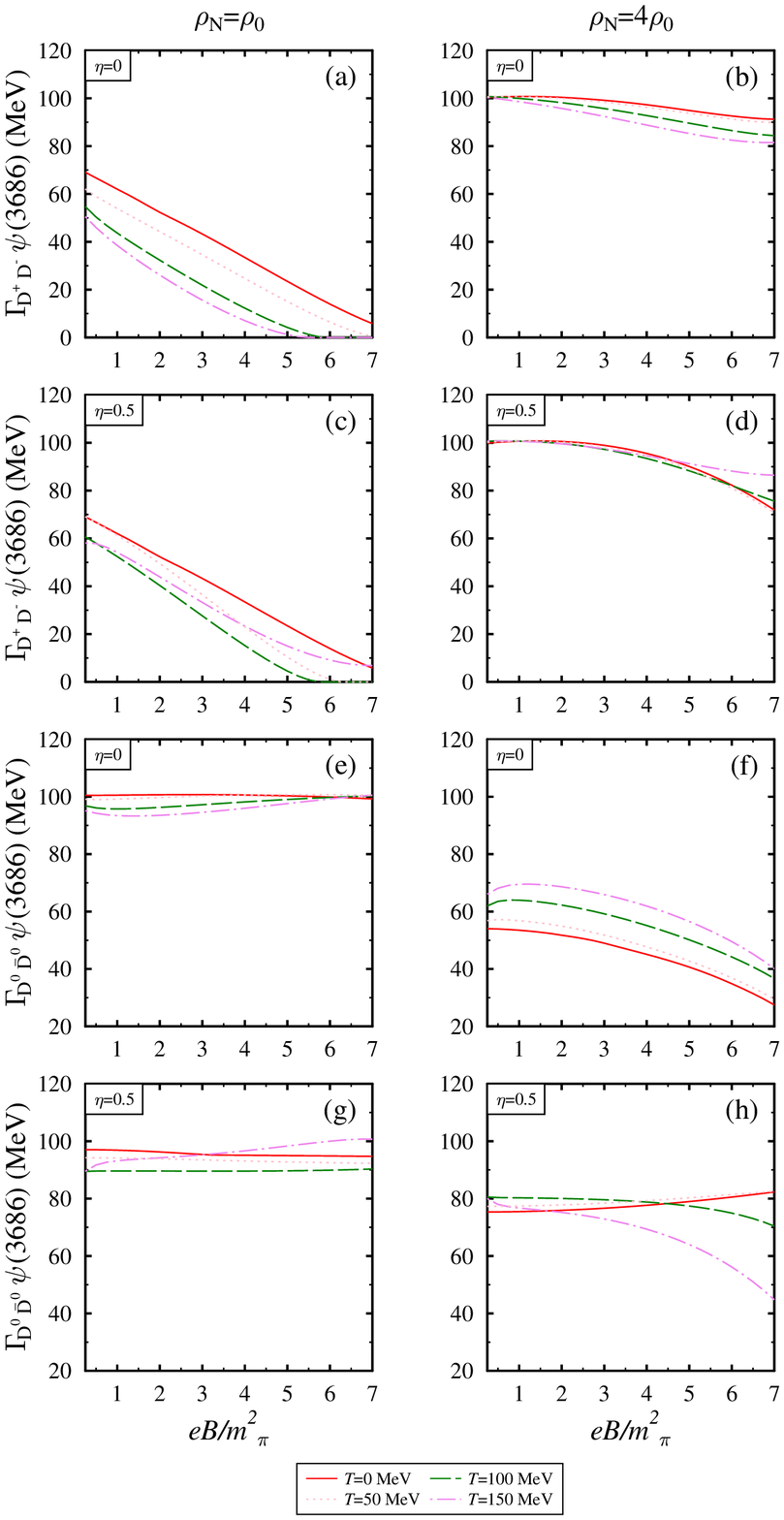}
\caption{(Color online) The effective decay width  of charmonium state, $\psi(3686)$, decaying into $D \bar D$ pairs is plotted as a function of magnetic field $eB$ under different conditions of medium. }
\label{dw1}
\end{figure}

\begin{figure}
\includegraphics[width=16cm,height=21cm]{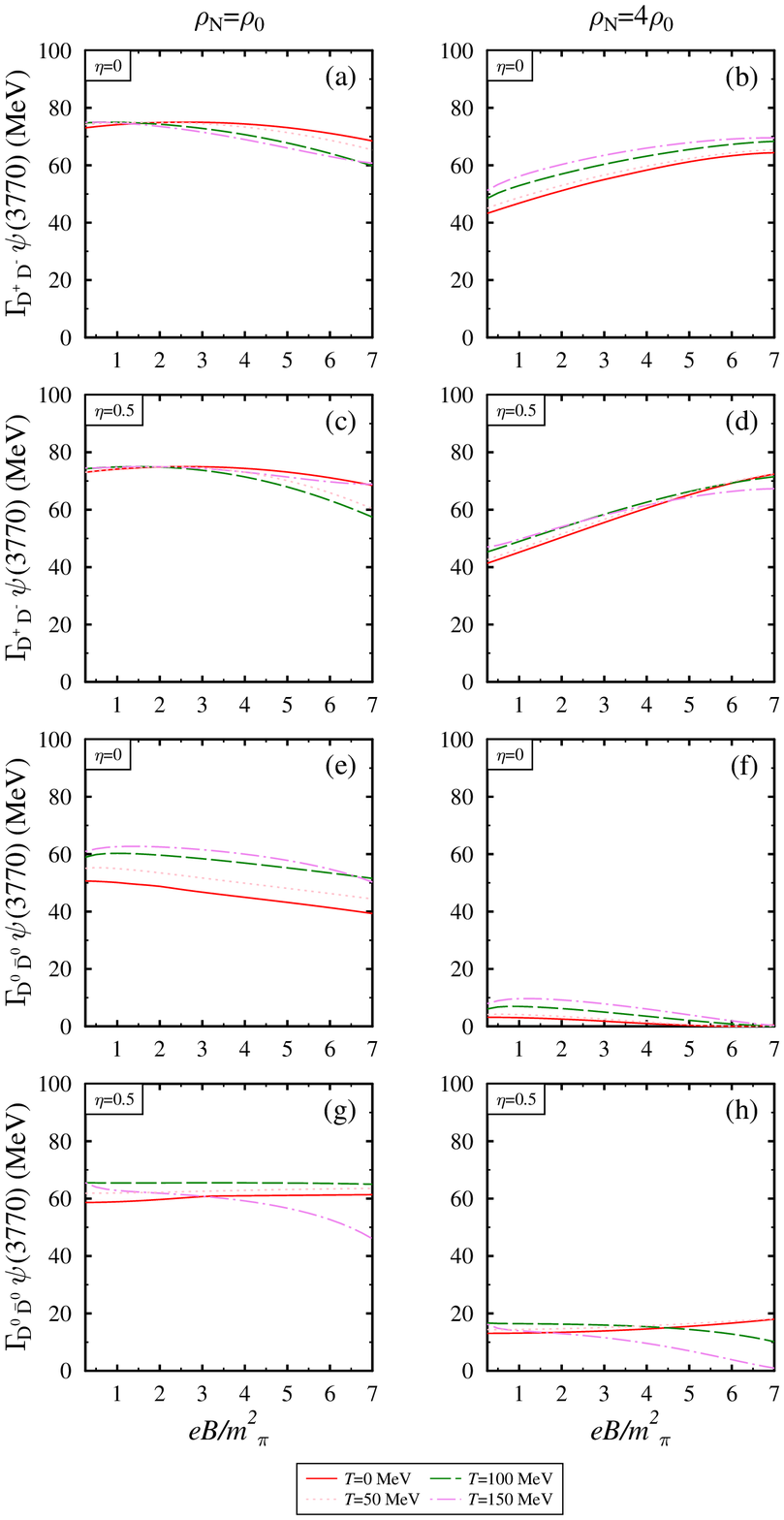}
\caption{(Color online)The effective decay width  of charmonium state, $\psi(3770)$, decaying into $D \bar D$ pairs is plotted as a function of magnetic field $eB$ under different conditions of medium.}
\label{dw2}
\end{figure}

Now, we will see how the obtained magnetically induced masses of pseudoscalar $D$ meson affect the in-medium decay width of higher charmonia ($\psi(3686),\psi(3770),{{\chi_c}_0}(3414),{{\chi_c}_2}(3556)$) decaying to $D \bar D$ pairs. We have neglected the medium modifications of the parent charmonia in the present work. In \cref{dw1} and \cref{dw2}, we have shown the variation of partial decay width of charmonia $\psi(3686)$ and $\psi(3770)$, respectively, decaying into $D \bar D$ mesons as a function of magnetic field in the nuclear matter. In Table \ref{tabledwdp}, we have listed the observed values of in-medium decay width of  charmonia decaying to $D^+D^-$ and $D^0 \bar D^0$  pairs.
As one can see from  \cref{dw1}, a non-zero value of decay width of $\psi(3686)$ to $D \bar D$ pair is observed because the threshold value of $D \bar D$ pair generation is less than the mass of parent meson. For decay channel $\psi(3686)\rightarrow D^+ D^-$, the value of partial decay width of $\psi(3686)$ decrease with the increase in magnetic field and it become zero for high value of magnetic field. The zero decay width arise when the threshold value of $D^+ D^-$ pair becomes more than the parent meson's mass.  The decrement in decay width  is due to the calculation of partial decay width (see Eq.(\ref{eqdw1})) which depends upon the energy $E_{D^+/ D^-}$ that further depends upon the in-medium mass $m^{**}_{D^+/ D^-}$. As discussed earlier, in the present work,  we have taken averaged meson current for $D^+$ and $D^-$ mesons (see Eq.\ref{mcq2}) and hence,  effective mass of $D^+$ and $D^-$ mesons will be same (similar will be the case for scalar $D^0$ and $\bar D^0$) \cite{Wang2015,Chhabra2017,Chhabra2018}. Since the effective mass of charged mesons increase in magnetic field and this will lead to increase of threshold value of $D^+$ $D^-$ pairs, and hence, a drop in dcay width. From \cref{ms_ps}, we can see the mass of $D^+$ mesons (threshold value of $D^+$ $D^-$ pairs) decrease with the increase in density whereas it increase with the increase in temperature.  This  explains why partial decay width of $\psi(3686)$ increases with the increase in density whereas decrease as a function of temperature.

 When we move from isospin symmetric medium to asymmetric medium, we observed  appreciable effects for density and magnetic field whereas least effects as a function of  temperature. This is the reflection of the observation of the in-medium mass of daughter meson in the respective medium, which occurs due to the unbalanced behaviour of scalar density of neutron and proton under the magnetic effects \cite{Kumar2019a}. In the decay channel $\psi(3686)\rightarrow D^0 \bar D^0$, in panel c (g) at $\eta$=0 (0.5), we observe that the decay width remains almost same (decrease) for low value of temperature whereas for higher value of temperature it increase as a function of magnetic field. Note that the trend of decay width of decay channel $\psi(3686)\rightarrow D^0 \bar D^0$ with magnetic field is opposite to decay channel $\psi(3686)\rightarrow D^+ D^-$. This is because being neutral charged mesons, $D^0$ and  $\bar D^0$ do not exhibit Landau quantization and hence mass does not increase but  decrease as a function of magnetic field as discussed in last subsection in detail.

In high-density regime, at $\eta$=0, the partial decay width of parent meson decrease with the increase in magnetic field and increases as a function of temperature. However, for $\eta$=0.5, it increases for lower value of temperature and decreases for higher value of temperature. These observations are contradictory to the explanation given in the previous paragraph as the high decrease in the mass cause decrease in decay width too. In Eq.(\ref{eqdw1}), the decay width is the product of Gaussian and polynomial expression. In high density, the polynomial part of the decay width dominates the Gaussian part.  In Table \ref{tabledwdp}, we have also compared the results with and without magnetic effects. We see that the increase in magnetic field causes more decrease in the value of decay width in low density as compared to high density for $D^+ D^-$ decay channel and for neutral $D$ meson pair, the decay width increases for low density and decreases for high density. 
 
 In \cref{dw2}, we plot the results of decay width  for decay channels $\psi(3770)\rightarrow D^+  D^-$ and $\psi(3770)\rightarrow D^0 \bar D^0$ for same parameters. For the former case, at lower density and particular value of temperature, the decay probability decreases slowly with the increase in magnetic field. The temperature and asymmetric effects are appreciable in the high magnetic field regime.  At high density, the trend of decay width is exactly opposite as a function of magnetic field and temperature.  The values of decay width increase with the increase in magnetic field which is due to the higher mass of parent meson $\psi(3770)$. The mass of parent meson rectifies the centre of mass momentum $p_D$, which results in the modification of the Gaussian expression. The interplay between Gaussian and polynomial expression leads to the above observations.  In the same figure, for decay channel $\psi(3770)\rightarrow D^0 \bar D^0$,  at $\eta$=0, the decay width decrease with the increasing magnetic field for high as well as low density. However, for high density the decay probability is very less due to the presence of Gaussian term ($e^{-\frac{y^2}{2(1+2r^2)}}$) in Eq(\ref{eqdw2})  and higher drop in mass of neutral $D^0$ meson. In highly asymmetric matter, the value of decay probability increase for low temperature and decrease for higher temperature as a function of magnetic field. This result is a reflection of the variation of in-medium mass in asymmetric magnetized nuclear matter. With the comparison of decay width of $\psi(3770)\rightarrow D^+ D^-$ in the absence of magnetic field, one can see that the value of decay width at particular combination of medium parameters increase in the magnetic field due to Landau levels and Gaussian interaction whereas it decreases for the $D^0 \bar D^0$  case. 
 
We have also calculated the decay widths of excited charmonium states $\chi_{c_0}$ and  $\chi_{c_2}$. The vacuum mass of $\chi_{c_0}$ and  $\chi_{c_2}$ is less than the threshold value for the  decay products ($D \bar D$)and therefore, the decay $\chi(3414)\rightarrow D \bar D$ and $\chi(3556)\rightarrow D \bar D$
is not possible. However, if the mass of $D$ meson drop appreciably and the threshold value become less than the mass of parent meson, then the decay is possible. We have observed zero decay probability for the $\chi(3414)$ mesons in $D^+ D^-$ pairs at all conditions of the medium. But for decay product $D^0 \bar D^0$, we have observed finite decay probability in high-density regime for non-magnetic case. For example, in non-magnetic cold nuclear matter at $\eta$=0 and $\rho_N$=$\rho_0$ $ (4\rho_0)$, the value of decay width of $\chi(3414)$ is 0 (5.6) MeV. Moreover, for decay channel  $\chi(3556)\rightarrow D \bar D$, we have observed a small finite decay width for low value of magnetic field up to  $eB/m^2_\pi$=2 at high density only. For instance, at $\eta=0$, $T$=0 and $4\rho_0$, we observed the values of decay width of $\chi(3556)$ to be 124 (158) MeV at $eB/m^2_\pi$=4 (6). The value of decay width decrease with the increase in temperature but the trend with respect to magnetic field remains same. Whereas for asymmetric matter it decreases for lower temperature and increases for higher temperature with respect to magnetic field. For example, at $\eta=0.5$, $\rho_N$=$4\rho_0$ and $T$=0 MeV, the value of decay width is 41 (35) MeV at $eB/m^2_\pi$=4 (6) whereas at $T$=100 MeV the value of decay width changes to 39 (47) MeV.  The value of decay width of both $\chi$ meson increase in the symmetric magnetic nuclear matter as compared to zero magnetic field data whereas for asymmetric nuclear matter it shows asymmetric variation due to $\delta$ field corrections. 

In  \cite{Mishra2019}, the magnetic field induced decay width of these four charmonium states have been calculated by considering  the in-medium masses of the charmonia as well as $D$ and  $\bar D$ mesons  in the combined approach of chiral SU(3) model and $^3P_0$ model at zero temperature only, whereas as discussed above in details we calculation are done at finite temperature which is important from heavy-ion collisions point of view.

\begin{table}
\begin{tabular}{|c|c|c|c|c|c|c|c|c|c|c|}
\hline
   & & \multicolumn{4}{c|}{$\eta$=0} & \multicolumn{4}{c|}{$\eta$=0.5}  \\
\cline{3-10}
& & \multicolumn{2}{c|}{T=0} & \multicolumn{2}{c|}{T=100 } &\multicolumn{2}{c|}{T=0} & \multicolumn{2}{c|}{T=100 }\\
\cline{3-10}
 &$eB/{{m}_{\pi}^2}$&  $\rho_0$ &$4\rho_0$&$\rho_0$&$4\rho_0$ &   $\rho_0$ & $4\rho_0 $ & $\rho_0$ & $4\rho_0$\\ \hline
$\Gamma_{D^+{D^-}}(\psi(3686))$ &0&75&97&59 &99.9 &82 &96 & 69  & 98\\  \cline{2-10} 
 & 4 &33 &97 & 12 & 93 &28 & 96 & 15 & 93 \\  \cline{2-10} 
 
 & 6 &14 &93 & 0 & 86 &4 & 82 & 0 & 82 \\ \cline{1-10}  

 $\Gamma_{D^+ {D^-}}(\psi(3770))$ &0 &71 &35 & 74 &41 &69 &32 &73 &36 \\   \cline{2-10} 
 & 4 &74 &58 & 71 & 63 &74 & 61 & 71 & 63 \\  \cline{2-10} 
 
 & 6 &71 &63 & 64 & 67 &68 & 69 & 63 & 69 \\ \cline{1-10}  
$\Gamma_{D^0{\bar D^0}}(\psi(3686))$ &0&100&47&94 &58 &95 &71 & 86  & 76\\  \cline{2-10} 
 & 4 &101 &45 & 98 & 55 &95 & 78 & 90 & 79 \\  \cline{2-10} 
 
 & 6 &100 &35 & 100 & 44 &95 & 81 & 90 & 75 \\ \cline{1-10}  

 $\Gamma_{D^0 {\bar D^0}}(\psi(3770))$ &0 &51 &1.3 & 62 &4.6 &60 &10 &67 &14 \\   \cline{2-10} 
 & 4 &45 &1 & 57 & 4 &61 & 15 & 65 & 15 \\  \cline{2-10} 
 
 & 6 &31 &0 & 53 & 1 &61 & 17 & 65 & 13 \\ \cline{1-10} 
 
\end{tabular}

\caption{In the above we list the values of in-medium decay width (in MeV) of higher  charmonium states $\psi(3686)$, $\psi(3770)$  to $D^+ {D^-}$ and $D^0 \bar D^0$ pairs  for different conditions of the medium.}
\label{tabledwdp}
\end{table}

%
%
%
%
%
%
%
%

\section{Conclusions}
\label{sec:4}

In the present investigation, we calculated the modification in the in-medium  masses and decay constant of pseudoscalar and scalar $D$ meson under the effect of external magnetic field at finite temperature, asymmetry and density of the nuclear medium. 
 To calculate the in-medium mass, we used the combined approach of QCD sum rules and chiral SU(3) model.   
In nuclear matter, the external magnetic field interacts with charged proton and uncharged neutron (due to non-zero anomalous magnetic moment) which results in the modification of scalar and vector density of nucleons. This magnetic field induced density is used to evaluate the coupled scalar fields of chiral model which are further used to calculate the medium modified scalar quark and gluon condensates \cite{Kumar2019a}. We found prominent effects of magnetic field on the charged $D^+$ and $D^+_0$ mesons. Whereas for neutral $D$ mesons, the effects are less. We found the negative (positive) mass shift for pseudoscalar (scalar) $D$ mesons. The temperature and density effects are also quite appreciable. The intervention of magnetic field depletes the effect of isospin asymmetry effects for the charged one but for uncharged mesons it was quite appreciable with crossover effects. As an application part, we calculated the in-medium decay width of higher charmonia in $^3P_0$ model and observed appreciable changes in the decay width of   $\psi(3686)$ and $\psi(3770)$ but less modification in $\chi(3414)$ and $\chi(3556)$. The calculated decay width may suppress the $J/\psi$ production and hence, may decrease its yield.
The experimental verification of obtained results can be done in the experimental facilities such as CBM, PANDA, J-PARC and NICA.

\begin{center}
\section*{Acknowledgement}
\end{center}

One of the author, (R.K)  sincerely acknowledge the support towards this work from Ministry of Science and Human Resources Development (MHRD), Government of India via Institute fellowship under National Institute of Technology Jalandhar.

\end{document}